\documentstyle[psfig,subfigure,graphicx,cite]{sup}
\vspace{-8pt}
\title[Superlattices and Microstructures, Vol. ??, No. ?, 1999]
{Quasiclassical Green's function approach to \\ mesoscopic
 superconductivity}
\author[Superlattices and Microstructures, Vol. ??, No. ?, 1999]{
Wolfgang Belzig$^1$, Frank K. Wilhelm$^1$, Christoph Bruder$^{1,2}$,\cr 
 Gerd Sch\"on$^1$, and Andrei D.Zaikin$^{1,3,4}$%
\cr\cr%
$^1${\normalsize\it 
Institut f\"ur Theoretische Festk\"orperphysik, Universit\"at
 Karlsruhe, D-76128 Karlsruhe, Germany}\cr
$^2${\normalsize\it 
Departement Physik, Universit\"at Basel, Klingelbergstrasse 82,
CH-4056 Basel, Switzerland}\cr
$^3$
{\normalsize\it Forschungszentrum Karlsruhe GmbH, Institut f\"ur Nanotechnologie,
D-76021 Karlsruhe, Germany}\cr
$^4$
{\normalsize\it P.N. Lebedev Physics Institute, Leninskii prospect 53,
117924 Moscow, Russia}
}

\pagerange{\pageref{firstpage}--\pageref{lastpage}}

\def\LaTeX{L\kern-.25em\raise.425ex\hbox{a}\kern-.075em\TeX}

\def\fakebold#1{\relax\ifvmode\leavevmode\fi%
\ifmmode%
\setbox0=\hbox{$#1$}%
\else%
\setbox0=\hbox{#1}%
\fi%
\kern-.02em\copy0 \kern-\wd0%
\kern .04em\copy0 \kern-\wd0%
\kern-.0125em\raise.02em\box0%
}%

\def\bbox#1{\mbox{\boldmath$#1$}}
\begin{document}
\label{firstpage}
\maketitle
\sloppy
\begin{center}
\received{(version of \today)}
\end{center}
%
\begin{abstract}
 Recent experiments on mesoscopic normal-metal--superconductor
 heterostructures resolve properties on length scales and at low 
 temperatures such that the temperature is below the Thouless energy
 $k_{\rm B}T \le E_{\rm Th}$. We describe the properties of these
 systems  within the framework of quasiclassical many-body
 techniques. Diffusive and ballistic systems are covered, both in
 equilibrium and nonequilibrium situations.  Thereby we demonstrate
 the common physical basis of various subtopics.
\end{abstract}

\section{Introduction}

Quasiclassical techniques \cite{eilenberger:68,larkin:68} have become a widely
used tool in the description of mesoscopic superconductivity.  In comparison
to early work which concentrated mostly on the regime near the superconducting
transition temperature, recent work resolved much smaller length scales and
covered lower temperatures.  At temperatures below the Thouless energy, i.e.
in dirty metals of size $d$ for $k_{\rm B}T \le E_{\rm Th} \sim \hbar D/d^2$,
novel features have been observed.  An important physical property is the
proximity effect in a normal metal in contact with a superconductor. It is
intimately related to the Andreev reflection \cite{andreev:64}, a process
in which an electron incident from a normal metal tries to enter a
superconductor.  If its energy lies below the gap, it can do so only if it
finds a partner electron with opposite momentum and spin to form a Cooper
pair, leaving a retroreflected hole in the normal metal. In this process two
charges are transferred into the superconductor. The reflection is
phase-coherent since the electron and the reflected hole combine in a way
consistent with the phase of the superconductor. The pair can maintain its
phase information in the normal metal over a distance $\sim \hbar v_{\rm
  F}/2E$, where $2E$ is the energy difference of the electron and hole.  This
space dependence, with diverging length scale at low energies, is responsible
for several of the mesoscopic effects to be discussed below.

The proximity effect is characterized by the existence of
superconducting correlations in the normal metal, 
expressed by a non-vanishing pair {\em amplitude\/}, $F \ne 0$. This
is to be distinguished from the pair  
{\em potential} $\Delta$, which in the absence of attractive
interactions vanishes in the normal metal. Finally, the
question whether the normal metal shows a {\em gap} in the
excitation spectrum is a more subtle issue. We will analyze below
how the proximity effect reduces the tunneling density of states
(DOS) at low energies, and, depending on the geometry, induces a gap or
a pseudo-gap. 

We will present here a unified view of both equilibrium and
nonequilibrium properties of normal-metal--superconductor proximity
systems. The basis are the quasiclassical Green's functions,
which will be reviewed in Section 2. Both the Matsubara
and the Keldysh formalism are introduced. We discuss the clean and
dirty limits, which allow systematic approximations. We finish
the section with remarks on the boundary conditions and the strategies of
solution. There exist several excellent introductions to this topic
\cite{Schmid,LO,rammer:86}. They have been written mostly with the
goal of describing massive superconductors at temperatures near the
transition temperature. Recent experimental work covered much lower
temperatures and resolved properties on sub-micron length scales. We,
therefore, present the theory and applications focusing on
nano-structured systems at low temperatures.

In the following sections we will discuss several applications of the
formalism and compare the results with recent experiments. In
the equilibrium case (Section 3) we study the change in the local
density of states (LDOS) in the vicinity of a normal metal -
superconductor interface. It shows a strong suppression
around the Fermi energy, leading to a mini-gap if the normal metal
is finite in size~\cite{golubov:88,belzig:96-2}. Our results agree with
experiments performed by the Saclay group~\cite{pothier:96}. Next, we
analyze the proximity induced magnetic screening in a normal metal in
contact with a superconductor. This allows us to describe the
experiments of Mota and coworkers~\cite{mota:89,belzig:98-3} on the
diamagnetic properties of normal - superconductor heterostructures. For
a comparison it is necessary to study the induced
superconductivity in a material with intermediate strength of the
impurity scattering. We found several new phenomena regarding the
non-locality 
of the current response in this regime\cite{belzig:98-2}. Finally we
consider the supercurrent through a normal metal sandwiched between
two superconductors, which allows us to interpret experimental results
of Courtois {\sl et al.}~\cite{CourtoisPRB}.

In Section 4 we study mesoscopic superconductors and heterostructures
in nonequilibrium situations. The quasiparticle distribution
function can be disturbed, e.g., by external applied gate voltages. Under
mesoscopic conditions, it acquires a specific double-step form, which
has been observed by the Saclay group~\cite{Pothier}. This
nonequilibrium form can be used, in a 4-terminal geometry, to tune a
supercurrent and, hence, to build a mesoscopic SNS transistor ~\cite{WSZ}.
Moreover, the SNS-junction can be switched by the gate voltages to
become a $\pi$-junction. We can explain in this way recent
experiments by the Groningen group \cite{Morpurgo,Baselmans,WeesHere}.
A different SN-heterostructure allows us to study the behavior of induced
superconducting correlations on a dissipative current. In systems with
transparent SN-contacts, the dissipative conductance is enhanced,
with a peculiar non-monotonic temperature- and
voltage-dependence, usually referred to as ``reentrant''
conductance~\cite{GWZ,NazStPRL}. This means,  at high temperatures the
conductance grows with decreasing $T$ since the proximity
effect develops, it reaches a a maximum at $k_{\rm B} T\approx E_{\rm Th}$
where $E_{\rm Th}$ is the Thouless energy, and finally it drops back
to the normal state value at $T=0$. This behavior has been observed in
experiments of the Grenoble group~ \cite{CourtoisLT,CourtoisHere}. In
more complex, multiply-connected structures the proximity conductance
can be modulated by an external magnetic flux. This modulation
is dominated by states close to the Fermi surface, for which
the spatial range of the proximity effect is large even at fairly
high temperatures.  This makes it possible to detect the influence of the 
proximity effect over distances much larger than the temperature
dependent coherence length~\cite{CourtoisPRL95,CourtoisHere,GWZ}.
In systems with tunneling barriers, the excess conductance competes
with the conductance suppression at the tunneling interfaces governed
by the opening of an induced gap. But even without tunneling barriers, the
conductance may be reduced as a consequence of a
4-terminal measurement in the thin film
geometry~\cite{PetrJETPL93,PetrPRL95,WZC}.

The purpose of this paper is to demonstrate by the use of quasiclassical
many-body techniques the common physical basis of several striking phenomena
observed in recent experiments on mesoscopic superconductors and
heterostructures. Of course, neither our presentation nor the bibliography 
can provide a complete review of the field. For a number of further
applications of quasiclassical techniques we refer to articles in Refs.
\cite{nato,curacao,Schoen,bruderhabil,raimondi,rainersauls} as well as the
articles of Nazarov, Volkov, and Yip in this volume.

\section{Quasiclassical formalism}

\subsection{Green's functions}
Quantum field theoretical methods in terms of Green's functions are
a powerful tool in all many-body problems
(see e.g. \cite{abrikosov:63,rickayzenmahan,Kadanoff}). 
Various systematic approximation methods
and computation schemes have been formulated for them. In this
section we will outline, how they can be used, within a
quasiclassical approximation to describe mesoscopic normal metal -
superconductor proximity systems.

The starting point for all problems in superconductivity is the Green's
function in Nambu space, which combines the particle and hole
space \cite{nambu:60}. Using the pseudo-spinors as a compact notation,
$\hat{\Psi}^\dagger=\left(\Psi_\uparrow^\dagger, \Psi_\downarrow\right)$,
we can express the time-ordered Green's function as
\begin{equation}
\hat{G}=-i\langle T\hat{\Psi} (1)\hat{\Psi}^\dagger (1^\prime)\rangle
=\left(\matrix{G(1,1^\prime)&F(1,1^\prime)\cr F^\dagger(1,1^\prime)&
G^\dagger(1,1^\prime)\cr} \right).
\label{GreensF}
\end{equation}
The Green's function contains an 'anomalous' component,
the pair amplitude $F=-i\langle T\Psi_\uparrow\Psi_\downarrow\rangle$,
characteristic for superconducting systems. Using the BCS Hamiltonian
we can write the Gorkov equation of motion~\cite{abrikosov:63} for
$\hat{G}$ as \footnote{Here and in the following we use units in which
$\hbar = k_B = 1$.}
\begin{eqnarray}
\left(\hat{G}_0^{-1}-\hat{\Delta}-\hat{\Sigma}_{imp}\right)(1,2)
\otimes\hat{G}(2,1^\prime)&=&\delta(1,1^\prime)
\nonumber\\ 
\hat{G}(1,2)\otimes \left(\hat{G}_0^{-1}
-\hat{\Delta}-\hat{\Sigma}_{imp}\right)(2,1^\prime)&=&\delta(1,1^\prime).
\label{Gorkov}
\end{eqnarray}
Here 1 and $1^\prime$ represent sets of space and time coordinates,
and $\otimes$ 
includes a convolution over the coordinates. 
The free Green's function in Nambu space reads
\begin{equation}
\hat{G}_0^{-1}(1,1^\prime)=\delta(1-1^\prime)\left[\hat{\tau}_3
\frac{\partial}{\partial{t_{1}}}
+\frac{1}{2m}\hat{\bbox\partial}_{\bbox r_1}^2-e\phi(1)+\mu\right]\;,
\end{equation}
where ${\hat{\bbox{\partial}}_{\bbox r}}
=\nabla_{\bbox{r}}-ie\bbox{A}(\bbox{r})\hat{\tau}_3$ is the
gauge-invariant spatial derivative. The part of the electron-phonon
self-energy which is responsible for superconductivity defines the
pair potential
\begin{equation}
 \hat{\Delta}(1,1^\prime)=\delta(1,1^\prime)
 \left(\matrix{0&\Delta\cr \Delta^\ast&0\cr}\right) \;\; \mbox{with} \;\;
 \Delta(1)=\lambda \lim_{2\longrightarrow 1+}F^\dagger(2,1)\ .
 \label{SCselfen}
\end{equation}
Here $\lambda$ is the strength of the attractive interaction. For a 
normal metal we have $\lambda=0$ and hence $\Delta=0$, but still the proximity 
effect manifests itself in a non-vanishing pair amplitude $F\not=0$. 
 
To simplify notations, we will from now on use center-of-mass coordinates in
space and time and Fourier-transform with respect to the relative coordinates.

Elastic impurity-scattering will be included within the framework
of the Born approximation. The impurity self-energy then reads
\begin{equation}
  \label{eq:impurity-selfenergy}
 \hat{\Sigma}(|p|,R,E,t)_{\rm imp} = 
 \frac{\pi}{2\tau}\left\langle\int d\xi\; 
\hat{G} (\bbox p,\bbox R,E,t) \right\rangle_{\bbox p_{\rm F}}\;,
\end{equation}
where $\tau$ is the elastic scattering time, $\xi=p^2/(2m)-\mu$ and $\langle
\dots \rangle _{p_{\rm F}}$ denotes averaging over the Fermi surface.  In
contrast, we can neglect inelastic scattering, since we will study mesoscopic
samples smaller than the inelastic length. Spin-flip scattering can be taken
into account by an expression similar to (\ref{eq:impurity-selfenergy}), see
\cite{Schmid,LO,rammer:86}.

The Gorkov equation can in principle be used to study mesoscopic proximity
systems. However, these are genuine inhomogeneous systems.  Hence, dealing
with full double-coordinate Green's functions -- while not impossible (see
e.g.\ \cite{Kulik,ishii:70}) -- may become very cumbersome, and typically one
makes use of quasiclassical approximations in the course of the calculations
anyhow.  For a systematic and more efficient approach it is an advantage to
perform this approximation already on the level of the equations of motion. \\

\subsection{The quasiclassical approximation}
We will now formulate the equations of motion within the quasiclassical
approximation. For simplicity, with the applications to be discussed in mind,
we restrict ourselves to equilibrium and stationary nonequilibrium situations,
although the quasiclassical approximation can be applied also to
time-dependent problems, e.g.\ relaxation processes and collective
modes~\cite{Schoen}.

The Green's function (\ref{GreensF}) oscillates as a function of the relative
coordinate $|\bbox{r}-\bbox{r^\prime}|$ on a scale of the Fermi wavelength
$\lambda_{\rm F}$. This is much shorter than the characteristic length scales
in the typical problems in superconductivity, $\xi_0= v_{\rm F}/\Delta$ and
$\xi_T= v_{\rm F}/2\pi T$. Moreover, in those problems, it is important to
study the phase of the {\em two-electron\/} wave function, which according to
the definition (\ref{GreensF}) depends on the center of mass coordinate. For
these reasons it is possible, and sufficient for most applications, to
integrate out the dependence on the relative coordinate. This has been
recognized first by Eilenberger \cite{eilenberger:68} and by Larkin and
Ovchinikov \cite{larkin:68}. We can add that the reduction is not allowed for
another class of mesoscopic effects, e.g. weak localization and persistent
currents, which are controlled by the phase-coherence of the {\em
  single-electron} wave function, contained in the relative coordinates of
(\ref{GreensF}). But these effects are usually much weaker than those related
to superconductivity. On the other hand, the reduction is possible also for
problems involving Andreev reflection, since the essential information is
again contained in the {\em difference} of electron and hole wave-vectors
close to the Fermi surface.

\subsubsection{Gradient expansion}
When integrating over  the difference variables  the convolution 
$\otimes$ in the Gorkov equation Eq.~(\ref{Gorkov}) (integration over 
internal variables) requires some care. It can be expressed, after the Fourier
transformation, as a Taylor series
\begin{equation}
 (A\otimes B) (\bbox p,\bbox r,E)=\exp\frac{i}{2}\left(
    \partial^{\rm A}_{\bbox r}\partial_{\bbox p}^B- 
  \partial^{\rm A}_{\bbox p}\partial^B_{\bbox r} \right) 
A(\bbox p,\bbox r,E)B(\bbox p,\bbox r,E)\;, 
\end{equation}
where $\bbox r$ refers to center of mass coordinates. 
In the problems to be discussed we can neglect short-range
oscillations, hence we expand this expression up to linear order. 

To proceed, we first subtract the Gorkov equation (\ref{Gorkov}) from its
conjugated form. We observe, that the Green's functions
and the self-energies are linear combinations of Pauli matrices {\em
not\/} including the unit matrix. This simplifies the equation of
motion to
\begin{equation}
 \left[i{\bbox{p\hat{\partial}_r}}
 +E\hat{\tau}_3-i\hat{\Delta}-\hat{\Sigma},\hat{G}\right]-
 \left\lbrace \bbox{\hat{\partial}_r}(\hat{\Delta}+\hat{\Sigma}+e\phi+\mu),
  \nabla_p\hat{G}\right\rbrace +\left\lbrace\nabla_p\hat{\Sigma},
\bbox{\hat{\partial}_r}\hat{G}\right\rbrace=0\; .
\label{GradEOM}
\end{equation}
This form is much simpler than the original. We note that it still
accounts for particle-hole 
asymmetry, which is necessary, e.g., for the description of
thermoelectric effects \cite{EckernSchmid,Chandra}. 

\subsubsection{Quasiclassical Green's functions}
On the one hand, we want to ignore the information contained in the fast
oscillations of the full Green's function $\hat{G}$ as a function of $|\bbox
r_1-\bbox r_2|$, which produces after Fourier transformation a pronounced peak
at $|\bbox p|=p_{\rm F}$. On the other hand, we have to pay attention to the
dependence on the transport direction, i.e. on the direction of the velocity,
$\bbox{v}_{\rm F}$, at the Fermi surface. To make this explicit we write
$\hat{G}(\xi,\bbox{v}_{\rm F},r,E)$, where $\xi=\frac{p^2}{2m}-\mu$ depends on
the magnitude of the momentum.  The quasiclassical Green's function is then
defined by
\begin{equation}
\hat{g}(\bbox{r},\bbox{v}_{\rm F},E) \equiv\frac{i}{\pi}\int
\begin{picture}(-10,5)
\put(-10,2){\line(1,0){7}}
\end{picture}
\begin{picture}(10,5)
\end{picture}
 d\xi\;
\hat{G}(\xi,\bbox{v_{\rm F}},\bbox{r},E)\;, 
\end{equation}
where the integration contour has two parts covering both half planes, see
\cite{Schmid} for details.
From Eq.~(\ref{GradEOM}) we get, from now on setting $p=p_{\rm F}$, the
Eilenberger equation of motion for quasiclassical Green's functions,
\begin{equation}
 -\left[\bbox{v}_{\rm F}\bbox{\hat{\partial}},
\hat{g}(\bbox r,\bbox{v}_{\rm F},E)\right]
 =\left[-iE\hat{\tau}_3+\hat{\Delta}
  +\frac{1}{2\tau}\langle\hat{g}(\bbox r,\bbox{v}_{\rm F},E)\rangle_{v_{\rm
 F}},\hat{g}(\bbox r,\bbox{v}_{\rm F},E)\right]
\; .
\label{eq:eilenberger}
\end{equation}
The elements of $\hat{g}$ are 
\begin{equation}
\hat{g}=\left(\matrix{g&f\cr f^\dagger&-g\cr}\right), 
\end{equation}
which is still a linear combination of three Pauli matrices
$\hat{\tau}_{1/2/3}$ with $f^\dagger$ being the time-reversed counterpart of
$f$.  This symmetry can be used for convenient parameterizations of the
Eilenberger equation (see Section \ref{ch:param}).

As the right-hand side of Eq.~(\ref{eq:eilenberger}) vanishes (in contrast to
the Gorkov equation (\ref{Gorkov})), it only defines the Green's function up
to a multiplicative constant. The constant can be fixed by the following
argument: As $\hat{g}$ is a linear combination of the Pauli matrices, the
square of the Green's function is proportional to the unit matrix
$\hat{g}\hat{g}=c \, \hat{1}$. From Eq.~(\ref{eq:eilenberger}) we see that the
proportionality constant $c$ is space-independent. If we now consider a system
containing a sufficiently large superconductor, we can identify a region
``deep inside the superconductor''. Here $\hat{g}$ equals its bulk value,
which can be calculated by performing the steps used in the quasiclassical
approximation for the known solution \cite{abrikosov:63} of the homogeneous
Gorkov equation Eq.~(\ref{Gorkov}). This procedure yields $c=1$, i.e.\ the
Green's functions are normalized
\begin{equation}
\hat{g}\hat{g}=\hat{1}, \quad \mbox{i.e.}  \quad \;\;\; g^2+ff^\dagger=1\; .
\label{norm}
\end{equation}
A more general, mathematical derivation of the normalization condition can
be found in \cite{EckernSchmid}\\

\subsection{Kinetics and time-ordering}
\label{ch:timeorder}
From the knowledge of the retarded and advanced Green's functions alone, we
can calculate energy-dependent quantities like the density of states
\begin{equation}
N(\bbox r,E)=N_0\hbox{Re}\left\{ \langle 
g^{\rm R}(\bbox r,\bbox{v_{\rm F}},E)\rangle \right\}.
\end{equation}
This, and similar expressions including the off-diagonal Green's function will
be denoted in the following as {\em spectral quantities}. In thermal
equilibrium these quantities also determine all properties of our system.
Equivalently, the thermal Green's function maybe calculated in Matsubara
imaginary-time technique. Then, the spectral quantities are found by analytic
continuation. 

In general, we need in addition information on how the quasiparticle states
are occupied, i.e. about distribution functions. To evaluate those under
nonequilibrium conditions we will use the Keldysh technique. Both techniques
are described in numerous references, so we will not rederive them but rather
explain the basic idea and provide a practical guideline for using them in
various situations.

\subsubsection{Matsubara technique}
\label{ch:matsubara}
The Matsubara Green's functions technique \cite{Matsubara} has been developed
to describe many-body systems in equilibrium at finite
temperature~\cite{abrikosov:63,rickayzenmahan}.  In thermal equilibrium, the
eigenvalues of physical observables do not depend on (real) time. In order to
calculate thermal expectation values, we trace over all states using the
Boltzmann weight $\exp(-H/T)$, which can be viewed as analytic continuation of
the time-evolution operator $\exp(iHt)$ to imaginary direction $\tau=it$. It
is sufficient to know this operator (and with it the Green's function) only in
the interval $0<\tau<1/T$. By Fourier transformation and exploiting the
Fermionic symmetry one sees that all the necessary information is contained in
the Green's functions defined for a discrete set of energies $E=i\omega_n$,
proportional to the Matsubara frequencies $\omega_n=(2n+1)\pi T$ with integer
values of $n$.

For those frequencies, in a bulk superconductor, the Green's functions read
\begin{equation}
g_{\omega}=\omega/\Omega\;\;\; , \; f_{\omega}=\Delta/\Omega \;\;\; 
\; \hbox{with} \; \Omega=\sqrt{|\Delta^2|+\omega^2}\label{Mat0}.
\end{equation}
This form will serve in the following frequently as a boundary condition.  The
Green's functions in imaginary times show usually no singular or oscillatory
structure and behave rather smooth and monotonic. This, together with the fact
that $g_{\omega}$ is real in the absence of fields or phase gradients,
simplifies the numerical calculations in thermal equilibrium.

Expectation values of physical quantities can be expressed via Green's
functions. To calculate thermal averages we rotate onto the imaginary time
axis and perform the quasiclassical approximation. For instance, the result
for the supercurrent density is
\begin{equation}
{\bbox j}({\bbox r})=-2ie\pi N_0T\sum_{\omega} \hbox{Tr} \langle
{\bbox v}_{\rm F} \hat{\tau}_3\hat{g}_\omega(v_{\rm F},r)\rangle\; .
\label{eq:matsubara_current}
\end{equation}
In general, it depends non-locally on the applied fields.  Similarly, the
self-consistency equation for the pair potential can be obtained from the
corresponding expression (\ref{SCselfen}) for 
Gorkov's Green's functions \cite{abrikosov:63}.
The integrations leading to the quasiclassical Green's function can be
performed, and the self-consistency relation reads
\begin{equation}
\Delta=\lambda N_02\pi T \sum_{\omega}\langle f_{\omega}\rangle.
\label{MatsubaraSelfcon}
\end{equation}

\subsubsection{The Keldysh technique}
The Keldysh Green's function technique~\cite{Keldysh} allows describing
many-body systems outside equilibrium. Pedagogical reviews are given in
articles \cite{rammer:86} and books \cite{Jensen}. It has proven particularly
useful in the description of nonequilibrium
superconductors~\cite{Schmid,LO,Schoen}.  Again, we will not rederive the
formalism, rather we summarize the concepts and some major results. We will
later restrict ourselves to the dirty limit, although the Keldysh technique can
be applied to ballistic systems as well \cite{Gunsenheimer}.

By means of the Keldysh technique it is possible to describe the real-time
evolution of systems in nonequilibrium and at finite temperature. We assume
that the system is initially, at $t=-\infty$, in a thermal equilibrium state.
The time evolution of the system, for instance of its density matrix, is
described by a forward and a backward propagator. Keldysh showed that this
real-time evolution, as well as an evolution along the imaginary time axis to
account for thermal averaging, can be combined to a propagation along a single
contour in the complex time plane, with a forward and backward branch running
parallel to the real axis and a vertical part between $t=-\infty-i/T$ and
$t=-\infty$ ~\cite{Keldysh}. Green's functions are defined on this contour
with time-ordering along the Keldysh contour. Depending on whether the time
arguments are on the forward or backward part, the Green's functions reduce to
different analytic parts. For instance, the time-ordered Green's functions or
the Kadanoff functions $G^{<(>)}$ can be obtained~ \cite{Kadanoff}. From those
one obtains by appropriate linear combinations the retarded and advanced
Green's functions $\hat{G}^{\rm R(A)}$ as well as the Keldysh Green's function
$G^{\rm K}$. They are related by
\begin{equation}
\hat{G}^>(1,1^\prime)=i\langle \Psi^\dagger(1^\prime)\Psi(1)\rangle
\;\;\; , \;\;
\hat{G}^<(1,1^\prime)=-i\langle \Psi(1)\Psi^\dagger(1^\prime)\rangle\; ,
\end{equation}
and
\begin{eqnarray}
\hat{G}^{\rm R}(1,1^\prime)&=&\theta(t_1-t_1^\prime)
\left[\hat{G}^<(1,1^\prime)-\hat{G}^>(1,1^\prime\right]
\nonumber\\
\hat{G}^{\rm A}(1,1^\prime)&=&-\theta(t_1^\prime-t_1)
\left[\hat{G}^<(1,1^\prime)-\hat{G}^>(1,1^\prime)\right]
\nonumber\\
\hat{G}^{\rm K}(1,1^\prime)&=&\hat{G}^>(1,1^\prime)+\hat{G}^<(1,1^\prime).
\end{eqnarray}
To compactify notations, these Green's functions can be written in matrix form
in `Keldysh space',
\begin{equation}
\check{G}=\left(\matrix{\hat{G}^{\rm R}&\hat{G}^{\rm K}\cr 0&
\hat{G}^{\rm A}\cr}\right).
\label{Keldyshmatrix}
\end{equation}
If we describe superconductors, each entry of this matrix is still a
$2\times2$ matrix in Nambu space.

The retarded and advanced Green's functions $G^{R(A)}$ determine what we will
call spectral, i.e.\ energy-dependent properties of the system. Their
Eilenberger equations have just the standard form, viz.,
Eq.~(\ref{eq:eilenberger}).  Usually they can be obtained by analytical
continuation of the Matsubara Green's function if we set
$\omega\longrightarrow -iE \pm0$.  On the other hand, $G^{\rm K}$ is needed to
account for properties of the system which depend on the nonequilibrium distribution
function. We will later relate it to this physical quantity and indicate ways
to calculate it.  Expectation values of physical quantities can again be
related to the Green's functions, now defined on the Keldysh contour
\cite{Schmid,LO,rammer:86,Kadanoff}. For instance the electrical current
becomes
\begin{equation}
{\bbox j}({\bbox r})=-eN_0\int_{-\infty}^\infty dE\;\hbox{Tr} \langle
{\bbox v}_{\rm F} \hat{\tau}_3
\hat{g}^{\rm K}({\bbox v}_{\rm F},{\bbox r},E)\rangle\; .
\label{eq:eilenkeldyshcurrent}
\end{equation}\\

\subsection{The dirty limit}
\subsubsection{Usadel equations in Matsubara technique}
Frequently the superconducting material has strong impurity scattering and is
described by the so-called `dirty limit'. The requirement is that the elastic
scattering self-energy dominates all other terms in the Eilenberger equation.
In this limit the electron motion is diffusive and the Green's functions are
nearly isotropic. I.e.\ one can expand the Green's functions in spherical
harmonics
\begin{equation}
 \label{eq:usadel_expansion}
 \hat{g}_\omega(\bbox{r},\bbox v_{\rm F})=\hat{G}_\omega(\bbox{r})+
 \bbox{v}_{\rm F}\hat{\bbox{g}}_\omega(\bbox{r})\; .
\end{equation}
Following the prevailing convention we denote the angular average of the
quasiclassical Green's function again by a capital letter. In the expansion we
assumed that $\bbox v_{\rm F}\hat{\bbox{g}}\ll\hat{G}$. From the normalization
condition (\ref{norm}) it follows that $\hat{G}_\omega^2(\bbox{r})=1$ and
$\{\hat{G}_\omega(\bbox{r}),\hat{\bbox{g}}_\omega(\bbox{r})\}=0$.  Angular
averaging of (\ref{eq:eilenberger}) yields
\begin{equation}
 \label{eq:usadel_isotropic}
 -\frac{1}{3}v^2_{\rm F}
 \left[\,{\bbox{\hat\partial}_{\bbox r}}\,,\,
  \hat{\bbox{g}}_\omega(\bbox{r})\,\right]=
 \left[\omega\hat{\tau}_3+\hat{\Delta}(\bbox{r})\,,
  \,\hat{G}_\omega(\bbox{r})\,\right]\; ,
\end{equation}
while averaging of (\ref{eq:eilenberger}) after multiplication by
$\bbox{v}_{\rm F}$, yields
\begin{equation}
 \label{eq:usadel_aniso}
 \hat{\bbox{g}}_\omega(\bbox{r})=
 -\tau\hat{G}_\omega(\bbox{r})\left[\,\hat{\bbox{\partial}}_{\bbox{r}}\, ,
  \,\hat{G}_\omega(\bbox{r})\,\right]\; .
\end{equation}
Here the condition $\hat g(\bbox r)/\tau \gg \omega\hat\tau_3+\hat\Delta$ has
been used. Inserting (\ref{eq:usadel_aniso}) into (\ref{eq:usadel_isotropic})
leads to the Usadel equation \cite{usadel:70},
\begin{equation}
 \label{eq:usadel_equation}
 D\left[\hat{\bbox{\partial}}_{\bbox r}\,,\,\hat{G}_\omega(\bbox{r})
  \left[\,\hat{\bbox{\partial}}_{\bbox{r}}\,,
   \,\hat{G}_\omega(\bbox{r})\,\right]\right]=
 \left[\omega\hat{\tau}_3+\hat{\Delta}(\bbox{r})\,,
  \,\hat{G}_\omega(\bbox{r})\,\right] \; ,
\end{equation}
where $D=v_{\rm F}^2\tau/3$ is the diffusion constant.
This equation is much simpler than the original Eilenberger equations and has
been widely applied to describe properties of
mesoscopic proximity systems. 

The current becomes in the dirty limit
\begin{equation}
 \label{eq:dirty_general_current}
 \bbox j (\bbox r)= -i\frac{\pi\sigma_{\rm N}}{2e}T\sum_\omega 
 \mbox{Tr}\hat\tau_3 \hat{G}_\omega(\bbox{r})
 \left[\,\hat{\bbox{\partial}}_{\bbox{r}}\,,\,\hat{G}_\omega(\bbox{r})\,\right]\;,
\end{equation}
where $\sigma_{\rm N}=2e^2N_0D$ is the normal-state conductivity. Note, that
a additional restrictions on the mean free path are
necessary for this local relation to hold. We will discuss this in
Section \ref{sec:diamag}.  E.g.\ for a bulk superconductor
with a real order parameter it is given by
\begin{equation}
 \label{eq:dirty_curr_field_relation}
 {\bbox j}({\bbox r})=-\pi\sigma_{\rm N} 
 T\sum_\omega \frac{\Delta^2}{\Omega^2}
 {\bbox A}({\bbox r}) \; ,
\end{equation}
which demonstrates an important property of the dirty limit, namely a
local supercurrent-field relation.\\ 

\subsubsection{The Usadel equation for Keldysh Green's functions}
\label{ch:kinetic}
The reduction to the dirty limit can be performed, similar to the
procedure outlined above, also for the Keldysh Green's functions.
The Usadel equation in Keldysh $\times$ Nambu space thus reads
\begin{equation}
 D\left[\bbox{\hat{\partial}_r}\,,\, 
  \check{G}(E,\bbox r)\left[{\bbox{\hat{\partial}_r}}\,,\,
   \check{G}(E,\bbox r)\right]\right] =
 \left[-iE\check\tau_3+\check\Delta , \check{G}(E,\bbox r)\right]\;,
\label{dirtyKeldysh}
\end{equation}
where 
\begin{equation}
\check{\tau}_3=\left(\matrix{\hat{\tau}_3&0\cr 0&\hat{\tau}_3\cr}\right)
\;\;\;
\check{\Delta}=\left(\matrix{\hat{\Delta}&0\cr 0&\hat{\Delta}\cr}\right).
\end{equation}

Although the Keldysh technique works also for
time-dependent situations \cite{Schmid,Schoen}, we restrict
ourselves here to stationary nonequilibrium problems.
We further consider only
structures in which all superconducting reservoirs are at
voltage $V=0$, thus avoiding effects related to the 
nonequilibrium Josephson effect. The normalization condition still holds in the form
$\check{G}\check{G}=\check{1}$. In terms of the components of
(\ref{Keldyshmatrix}) it implies
\begin{equation} 
\hat{G}^{R}\hat{G}^{\rm R}=\hat{G}^{\rm A}\hat{G}^{\rm A}=\hat{1} \; ,
\;\;\; \mbox{and} \;\;\;
\hat{G}^{\rm R}\hat{G}^{\rm K}+\hat{G}^{\rm K}\hat{G}^{\rm A}=0\; .
\end{equation}
As a consequence of the second relation $G^{\rm K}$ can be parameterized as
\begin{equation}
\hat{G}^{\rm K}=\hat{G}^{\rm R}\hat{h}-\hat{h}\hat{G}^{\rm A} \; .
\label{ParamKel}
\end{equation}
From the Keldysh component (i.e.\ upper right) of the
Keldysh-Usadel equation (\ref{dirtyKeldysh}) we obtain a kinetic equation
of motion for the distribution matrix $\hat{h}$
\begin{eqnarray}
D\left[\nabla^2\hat{h}+(\hat{G}^{\rm R}\nabla\hat{G}^{\rm R})\nabla\hat{h}-
\nabla\hat{h}(\hat{G}^{\rm A}\nabla\hat{G}^{\rm A}) 
-\nabla(\hat{G}^{\rm R}(\nabla\hat{h})\hat{G}^{\rm A})\right]
&&
\nonumber\\
-\left(\hat{G}^{\rm R}[\hat{\Delta},\hat{h}]
-[\hat{\Delta},\hat{h}]\hat{G}^{\rm A}\right)
+iE\left(\hat{G}^{\rm R}[\hat{h},\hat{\tau}_3]-[\hat{h},\hat{\tau}_3]
\hat{G}^{\rm A}\right)&=&0\; .
\end{eqnarray}
Here, we made use of the fact that $G^{\rm R(A)}$ satisfy the respective
components of the Usadel equation.  The kinetic equation has only two
independent entries. Since, furthermore, it is a linear equation, we can
assume $\hat{h}$ to be diagonal. Returning to the definitions, we can relate
it to the distribution functions for electrons and holes
\begin{equation}
\hat{h}=\left(\matrix{1-2f_{el}&0\cr 0&2f_h-1}\right) \; ,
\end{equation}
where the energy is measured from the chemical potential of the superconductor.
In thermal equilibrium, for instance in the reservoirs at voltage
$V_R$,  it can be expressed by Fermi functions 
\begin{equation}
\hat{h}_{eq}=\left(\matrix{1-2f(E)&0\cr 0&2f(-E)-1}\right)
=\left(\matrix{\tanh\left(\frac{E+eV_R}{2T}\right)&0\cr 
0&\tanh\left(\frac{E-eV_R}{2T}\right)\cr}\right).
\end{equation}
For practical calculations, it is convenient to split the distribution
matrix into odd and even components with respect to the Fermi surface
$\hat{h}=f_{\rm L}+f_{\rm T}\hat{\tau}_3$.  The notation $f_{\rm L/T}$
refers to `longitudinal' and `transverse' changes of the order
parameter associated with the respective nonequilibrium
distributions~\cite{SchmidSchoen,Tinkham}. Here it is sufficient to
mention that deviations of $f_{\rm L}$ from equilibrium are related to
an effective temperature change, whereas $f_{\rm T}$ is related to an
effective chemical potential shift.  The equilibrium forms in the
reservoirs are
\begin{equation}
 f_{\rm L(T)} = \frac12 \left[\tanh\left(\frac{E+V_R}{2T}\right)
  +(-)\tanh\left(\frac{E-V_R}{2T}\right)\right]
 \label{KineticBoundary}.
\end{equation}
We can recover the usual electron distribution
function in the end of the calculation as
$2f(E)=1-f_{\rm L}(E)-f_{\rm T}(E)$. 

The kinetic equations for the two components  reduce to two
coupled diffusion equations
\begin{eqnarray}
 \nabla({\cal D}_{\rm T}\nabla f_{\rm T})+2\hbox{Im}\{j_E\}\nabla f_{\rm L}&=&
 2 \frac{{\cal R}}{D}\, f_{\rm T}
\label{kinetic1}\\
 \nabla({\cal D}_{\rm L}\nabla f_{\rm L})+2\hbox{Im}\{j_E\}\nabla f_{\rm T}
 &=&0\label{kinetic2}
\end{eqnarray}
with energy dependent spectral quantities 
\begin{eqnarray}
 \hbox{2Im}\{j_E\}&=&\frac14\hbox{Tr}\left[\hat{\tau}_3\left(\hat{G}^{\rm R}\hat{\nabla}\hat{G}^{\rm R}-
\hat{G}^{\rm A}\hat{\nabla}\hat{G}^{\rm A}\right)\right]\; ,\\
{\cal D}_{\rm T}&=&\frac14\hbox{Tr}\left(1-\hat{G}^{\rm R}\hat{\tau}_3\hat{G}^{\rm A}\hat{\tau}_3\right)
\label{DiffusionT}\; ,\\
{\cal D}_{\rm L}&=&\frac14\hbox{Tr}\left(1-\hat{G}^{\rm R}\hat{G}^{\rm A}\right)\; ,\\
{\cal R}&=&\frac14\hbox{Tr}\left[\hat{\Delta}\left(\hat{G}^{\rm R}+
\hat{G}^{\rm A}
\right)\right]\; .
\label{diffkoeff}
\end{eqnarray}
Note, that the diffusion constant is generalized to energy
dependent spectral diffusion coefficients, ${\cal D}_{\rm T/L}(E)$. They 
are even functions of energy, whereas the the spectral supercurrent,
 $j_E$, is an odd function of the energy. Recently, it has been argued, that in 
some particular situations, the diffusion equation should be supplemented by
another term usually neglected\cite{Galak}.

The right-hand side of eq. (\ref{kinetic1}) implies, that $f_{\rm T}$
relaxes to 0 inside the superconductor on 
a scale of $\xi_0$, i.e. it adjusts the chemical potential to that of
the superconductor. In contrast the right-hand-side of Eq. (\ref{kinetic2})
 vanishes. This component $f_{\rm L}$ relaxes only due 
to inelastic processes, which are neglected here, on a longer length scale
 \cite{SchmidSchoen}.

\subsubsection{Physical quantities and Keldysh Green's functions}

We now list some results for physical quantities expressed by Keldysh-Usadel
Green's functions.
The self-consistency equation for the order parameter reads in analogy
to (\ref{MatsubaraSelfcon})
\begin{equation}
\Delta=\frac{\lambda}{4i}\int_{-\infty}^\infty dE\; f_{\rm L} 
(F^{\rm R}-F^{\rm A})\; ,
\label{KeldyshSelfcon}
\end{equation}
while the current becomes
\begin{equation}
j=\frac{\sigma_{\rm N}}{2}\int dE \hbox{Tr}\left[\hat{\tau}_3(
\hat{G}^{\rm R}\nabla\hat{G}^{\rm K}+\hat{G}^{\rm K}\nabla\hat{G}^{\rm
A} )\right] \, .
\end{equation}
We can now use the parameterization of the Keldysh Green's function by
distribution functions, Eq.~(\ref{ParamKel}), and obtain terms
proportional to the two distribution functions and their respective
derivatives. By studying the parity of the terms  in energy we note
that some terms vanish, and we end up with the supercurrent
\begin{equation}
 j_{\rm s}=\frac{\sigma_{\rm N}}{2}\int_{-\infty}^\infty dE\;
 f_{\rm L}(E) \, \hbox{Im}\{j_E\}
\label{KeldyshSupercurrent}
\end{equation}
and the dissipative current component
\begin{equation}
 j_{\rm n}=
 \frac{\sigma_{\rm N}}{2}\int_{-\infty}^\infty dE\; 
 {\cal D}_{\rm T}(E) \, \nabla f_{\rm T}(E) \; .
\label{Dissip}
\end{equation}
The kinetic equation (\ref{kinetic1}) guarantees the
conservation of the total current, $j=j_s+j_n$, in the normal metal. The
other kinetic equation (\ref{kinetic2}) can be identified as
describing the convective part of the thermal current.\\

\subsection{Boundary Conditions}
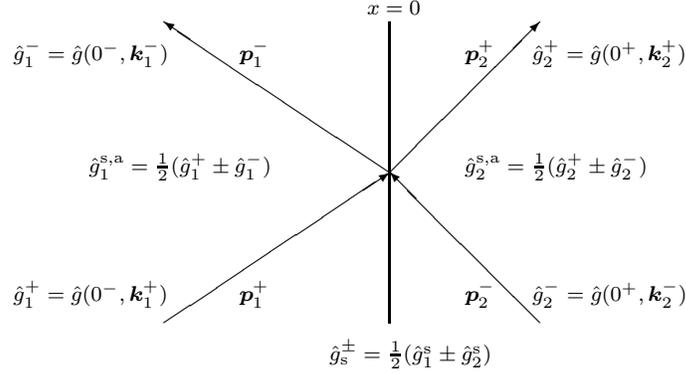
\begin{figure}[htbp]
\begin{center}
 \unitlength1cm
\begin{picture}(10,6)
 {\thicklines\put(5,1){\line(0,1){4}}} \put(2,1){\vector(3,2){3}}
 \put(5,3){\vector(-3,2){3}} \put(5,3){\vector(1,1){2}}
 \put(7,1){\vector(-1,1){2}} {\thicklines\put(5,2){\line(0,1){2}}}
 \put(3.0,1.3){$\bbox p_1^+$} 
 \put(3.0,4.5){$\bbox p_1^-$}
 \put(6.0,4.5){$\bbox p_2^+$} 
 \put(6.0,1.3){$\bbox p_2^-$}
 \put(0,1.3){$\hat g_1^+=\hat g(0^-,\bbox k_1^+)$} 
 \put(0,4.5){$\hat g_1^-=\hat g(0^-,\bbox k_1^-)$}
 \put(6.9,1.3){$\hat g_2^-=\hat g(0^+,\bbox k_2^-)$} 
 \put(6.9,4.5){$\hat g_2^+=\hat g(0^+,\bbox k_2^+)$}
 \put(6,3){$\hat g_2^{\rm s,a}=\frac 12(\hat g_2^+\pm\hat g_2^-)$}
 \put(1,3){$\hat g_1^{\rm s,a}=\frac 12(\hat g_1^+\pm\hat g_1^-)$}
 \put(4.2,0.5){$\hat g_{\rm s}^\pm=\frac 12(\hat g_1^{\rm
 s}\pm\hat g_2^{\rm s})$}
 \put(4.7,5.1){$x=0$}
\end{picture}
\caption{Trajectories at an interface. The parallel momenta are conserved,
   but for different Fermi velocities on both sides the perpendicular
   momenta differ. The Green's functions on the trajectories involved
   are indicated by $\hat g_{1,2}^{+,-}$.}
 \label{fig:interface_momenta}
\end{center}
\end{figure}

In the quasiclassical approximation the information on length scales of the
order of the Fermi wavelength have been integrated out. Consequently one can
no longer account directly for the effect of potential barriers or interfaces
on this level. It turns out, however, from a study of the full theory, that
potential barriers and interfaces can be accounted for by effective boundary
conditions for the quasiclassical Green's functions.  A derivation of boundary
conditions valid at arbitrary transmission of the interface has been given by
Zaitsev \cite{zaitsev:84}. These boundary conditions couple the classically
transmitted and reflected trajectories.  The corresponding wave vectors $\bbox
k_{1,2}^\pm$ are shown in Fig.~\ref{fig:interface_momenta}.
The in- and outgoing Green's functions $\hat g_{1,2}^\pm$ on both sides
(1 and 2) on these trajectories are combined as
indicated in Fig.~\ref{fig:interface_momenta}. They fulfill the boundary
conditions
\begin{eqnarray}
 \label{eq:zaitsev_conditions}
 \hat g_1^{\rm a} & = & \hat g_2^{\rm a} \; ,
\nonumber\\
 \hat g^{\rm a}\left[ R (\hat g_{\rm s}^+)^2+(g_{\rm s}^-)^2\right] & = &
 T \hat g_{\rm s}^-\hat g_{\rm s}^+\; ,
\end{eqnarray}
where $R=1-T$ is the momentum-dependent reflection coefficient. The functions
$\hat g_{1,2}^{\rm a}$ and $\hat g_{\rm s}^{+,-}$ are defined in
Fig.~\ref{fig:interface_momenta}. Note, that current
conservation through the boundary is ensured by the first of
these equations. The boundary conditions are also valid 
for Keldysh-Green's function, provided the Nambu matrices are replaced by
Keldysh matrices.

The boundary conditions can be simplified in the dirty limit for low
transparencies of the interfaces. As was shown by Kuprianov and
Lukichev \cite{kuprianov:88} they read in this limit
\begin{eqnarray}
 \label{eq:kuprianov_conditions1}
 p_{F1}^2l_1\hat G_1\frac{d}{dx}\hat G_1 & = &
 p_{F2}^2l_2\hat G_2\frac{d}{dx}\hat G_2\; , 
\nonumber\\
 \label{eq:kuprianov_conditions2}
 l_2\hat G_2\frac{d}{dx} G_2 & = &
 t\left[\hat G_2,\hat G_1\right]\; .
\end{eqnarray}
Here $t=\langle p_{{\rm F}2x}T/p_{{\rm F}2}R\rangle$ is a (small) parameter
related to the transparency, and $p_{{\rm F}2x}$ is the projection of the
Fermi momentum onto the normal of the interface. Recently Lambert {\it et al.}
\cite{lambert:96} showed that the second of these conditions constitutes the
first term of an expansion in the parameter $t$, and they
calculated the second term in this expansion.\\

\subsection{Parameterizations}
\label{ch:param}
For the further handling of the quasiclassical equations, both numerically and 
analytically, two parameterizations have turned out
to be especially useful: the
Riccati- and the $\theta$-parameterization.

\subsubsection{Riccati parameterization}
We will make use of the Riccati parameterization mostly in equilibrium
problems for general strength of the impurity scattering. In this case
we write the Matsubara Green's functions as \cite{schopohl}
\begin{equation}
 \label{eq:riccati_parametrization}
 \hat{g}_\omega(\bbox r,\bbox{v}_{\rm F})=\frac{1}{1+aa^\dagger}
 \left(
  \begin{array}[c]{cc}
   1-aa^\dagger & 2a\\
   2a^\dagger & -1+aa^\dagger
  \end{array}
 \right)\; , 
\end{equation}
where the functions $a_\omega(\bbox r,\bbox{v}_{\rm F})$ 
and $a^\dagger_\omega(\bbox r,\bbox{v}_{\rm F})$ obey the equations
\begin{eqnarray}
 \label{eq:riccati_equations}
 -\bbox{v}_{\rm F}\bbox{\nabla}a & = & 2\tilde{\omega}a+
 \tilde{\Delta}^* a^2 -\tilde{\Delta}\; ,
\nonumber\\
 \bbox{v}_{\rm F}\bbox{\nabla}a^\dagger & = & 
 2\tilde{\omega}a^\dagger+\tilde{\Delta} a^{\dagger2} -\tilde{\Delta}^*\; .
\end{eqnarray}
Here 
\begin{equation}
  \tilde\omega=\omega+\langle g\rangle/2\tau \; \; \mbox{and} \; \;
  \tilde\Delta=\Delta + \langle f \rangle / 2\tau
  \label{tildes}
\end{equation}
are the renormalized energy and pair potential. These equations are of the
Riccati type. The functions $a$ and $a^\dagger$ appearing in this
parameterization are directly related to the coefficients of the Andreev
amplitudes. This can be seen, by noting that (\ref{eq:riccati_equations}) in
the clean limit directly follow from the Andreev equations \cite{andreev:64}
identifying $a=u/v$\cite{schopohl:98}.
For numerical purposes these equations are well suited, since they
are two {\em uncoupled stable} differential equations (one for each direction
of integration). This should be contrasted to the original Eilenberger
equations, which are coupled and unstable differential equations. These
equations also provide a basis to treat the linear response to an external
field \cite{belzig:98-2,eschrig:97}.

\subsubsection{$\theta$ parameterization}
The dirty-limit equations, which do not dependent on 
$\bbox{v}_{\rm F}$ any more, allow a simpler parameterization 
\begin{eqnarray}
 \label{eq:theta_parametrization}
 \hat{G}^{\rm R}(E,\bbox r)&=&
 \left(
  \begin{array}[c]{cc}
   \cosh(\theta) & 
   \sinh(\theta)\exp(i\chi)\\
   -\sinh(\theta)\exp(-i\chi) & 
   -\cosh(\theta)
  \end{array}
 \right)
\nonumber\\
 \hat{G}^{\rm A}(E,\bbox r)&=&
 \left(
  \begin{array}[c]{cc}
   -\cosh(\bar{\theta}) & 
   \sinh(\bar{\theta})\exp(i\bar{\chi})\\
   -\sinh(\bar{\theta})\exp(-i\bar{\chi}) & 
   \cosh(\bar\theta)
  \end{array}
 \right)\; . 
\end{eqnarray}
Here, $\theta(E, \bbox r)$ and $\chi(E,\bbox r)$ are complex
functions the bar denotes complex conjugation
\footnote{It is also possible to parameterize 
using $\sin$- and $\cos$-functions. We prefer the present
convention, which requires to choose the gauge of $\Delta$ such that
for a real pair potential, $\Delta=|\Delta|$, we
have $\hat{\Delta}\propto\tau_2$. Compared to our previous
definition in eq. (\ref{SCselfen}) we have to shift the
phase of the pair potential, $\phi$, by a gauge transformation of the
order parameter phase by $\pi/2$.}. 
The Usadel equations in the normal metal are now written as
\begin{eqnarray}
D\partial^2_x\theta &=&-2iE\sinh\theta +\frac{D}{2}
\left(\partial_x\chi \right)^2\sinh2\theta \nonumber\\
\partial_xj_E&=&0\;\;\; \mbox{,} \;\;\;
j_E=2\sinh^2\theta \partial_x\chi\; . 
\label{retard}
\end{eqnarray} 
In imaginary time, it is more convenient 
to parameterize \cite{ZaikinZharkov}
\begin{equation}
 \label{eq:theta_parametrization_Matsubara}
 \hat g_\omega(\bbox r)=
 \left(
  \begin{array}[c]{cc}
   \cos(\theta) & 
   \sin(\theta)\exp(i\chi)\\
   \sin(\theta)\exp(-i\chi) & 
   -\cos(\theta)
  \end{array}
 \right)\; , 
\end{equation}
where $\theta(\omega,\bbox r)$ and $\chi(\omega,\bbox r)$ 
are now real functions. 
At the interface to a
superconducting reservoir with pair amplitude $\Delta_0e^{i\phi_0}$ (which
vanishes in normal conductors), which is connected to the probe by good
metallic contacts, the boundary conditions read
\begin{equation}
  \begin{array}[c]{rclcl}
    \theta&=&\theta_{\rm S}&=&\hbox{Arctanh}(\Delta/E)\;,\\
    \chi&=&\phi_0\; .
  \end{array}
\label{BCUsa}
\end{equation}
For a normal reservoir, we set $\theta=0$.  In the presence of tunneling
barriers, the boundary condition get modified. For instance, if we
consider a normal layer
with resistance per length $\tilde{R}_{\rm N}$ coupled via a barrier with
resistance $R_{\rm t}$ to a superconductor, the condition  on $\theta$
following from Eq.~(\ref{eq:kuprianov_conditions1}) reads
\begin{equation}
r_{\rm t}\,\partial_x\theta=\sinh(\theta-\theta_{\rm S})\; .
\end{equation}
The direction of the derivative points away from the
superconductor described by $\theta_S$, and $r_{\rm
 t}=R_{\rm t}/\tilde{R}_{\rm N}$ is the ratio of the resistances. For the
kinetic equation Eq.~(\ref{kinetic2}), the boundary 
condition is
\begin{equation}
r_{\rm t} \,D\partial_x f_{\rm T/L}=M_{\rm T/L}(f_{T\rm 0/L0}-f_{\rm T/L})\;,
\label{GolKupBC}
\end{equation}
where $f_{T0/L0}$ are the components of the reservoir distribution function
(\ref{KineticBoundary}). Using the Green's functions $G^{\rm R}_S$ and
$F^{\rm R}_S$  of the superconductor we defined  the spectral functions 
\begin{equation}M_{\rm T/L}=\hbox{Re}(G^{\rm R}_S)\hbox{Re}(G^{\rm R})
\pm\hbox{Im}(F^{\rm R}_S)\hbox{Im}({F^\dagger}^{\rm R})\; .
\end{equation} \\

\subsection{Solution strategies}
In some cases we can derive approximate analytical solutions, for instance
linearize the equation (see e.g. \cite{volkov:here}). In general, however, we
have to rely on numerical solutions.  To find the solution for a given
transport problem, we can proceed as follows:
\begin{enumerate}
\item{Start with a given $\Delta({\bbox r})$}.
\item{Solve the retarded Usadel equation}.
\item{Take the solutions to calculate $N(E)$, $j_E$, and ${\cal D}_{L/T}$}.
\item{Solve the kinetic equations}.
\item{Calculate a new $\Delta({\bbox r})$, and iterate until self-consistency 
is achieved}.
\end{enumerate}
It is worth noting that considerable simplifications arise if one is only
interested in equilibrium properties. Then it is possible to determine the
self-consistent problem in the Matsubara representation, which is numerically
much more stable than the real-time version.  (In the latter one encounters
frequently the cancelation of large terms.)  Even if one is interested in
transport properties it is of advantage to proceed in two steps. First, the
pair potential, which does not depend on energy, is determined in the
Matsubara representation. Then, after self-consistency has been achieved, one
proceeds to solve the real-time/ real-energy quasiclassical equations to
obtain spectral
quantities and the transport properties.\\

\section{Spectral quantities and equilibrium properties}
\subsection{Density of states}
\label{ch:DOS}
As a first application we study of the local density of states in spatially
inhomogeneous dirty systems. Examples are a normal metal in contact with a
superconductor, inhomogeneous superconductors or boundaries of superconductors
with unconventional symmetry. Depending on the problem, the pair potential
$\Delta({\bbox r})$ has to be determined self-consistently. As mentioned
above, for the ease of the numerical procedure this is best performed in
imaginary frequencies. The next task is to solve the Usadel equation for real
energies for a given pair potential. Once we have determined the matrix
Green's function $\hat G^{\rm R}(E,\bbox r)$, the local density of states is
given by
\begin{equation}
 \label{eq:general_dos}
  N(E,\bbox r)= N_0 \mbox{Re}\left[\hat G^{\rm R}(E,\bbox r)\right]= 
  N_0\mbox{Re}\left[\cosh(\theta(E,\bbox r)\right] \; ,
\end{equation}
and similar for other spectral functions.

In this chapter we will calculate the local density of states $N(E,\bbox r)$
for geometries in which a normal metal and a superconductor are in
contact\cite{belzig:96-2}. First we consider the simplest case of an infinite
normal metal in contact with an infinite superconductor. In this system the
spatially resolved density of states has recently been measured with the help
of tunnel junctions contacting the normal metal in different positions
\cite{pothier:96}.  As a second example, we consider a finite thickness normal
metal in contact with a superconductor. In this case, there is a gap in the
excitation spectrum of the normal metal. The form of the density of states in
this system leads to rather unusual current-voltage characteristics of SNINS
tunnel junctions. This has recently been measured in proximity point contacts
\cite{scheer:98}, in which the `insulator' is an atomic point contact with
variable transparency \cite{scheer:97}.  The final system examined is an
overlap junction, where the normal metal and the superconductor have an finite
overlap. Interestingly, in this system a local gap-like feature in the
overlap region exists. In all examples in this section the pair potential is
calculated self-consistently.

\subsubsection{Infinite proximity system}
\begin{figure}[ht]
  \begin{center}
    \includegraphics[width=8cm]{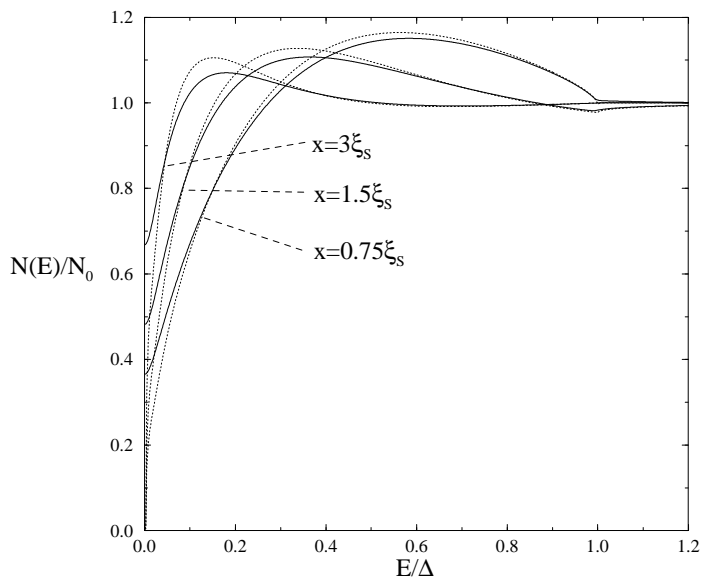}
    \caption[]{
      Density of states on the normal side of an N-S boundary for two spin-flip
      scattering rates: $\Gamma_{{\rm sf}}=0$ (dotted lines) and $\Gamma_{{\rm
          sf}}=0.015\Delta$ (solid lines).}
    \label{fig:dos.inf.n}
  \end{center}
\end{figure}
We consider a quasi one-dimensional geometry, i.~e., an infinite normal metal
is in contact with a superconductor of the same cross section.  The solution
of the Usadel equations yields the DOS on the normal side at different
distances from the NS-boundary as shown in Fig.~\ref{fig:dos.inf.n}. The peak
height and position change with distance. In the absence of pair-breaking
effects the DOS vanishes at the Fermi level for all distances (dotted curves).
Inclusion of a pair-breaking mechanism, $\Gamma_{\rm sf} \ne 0$, (solid
curves) regularizes the DOS at the Fermi level, and also suppresses the peak
height. The curves are in qualitative agreement with experimental data
presented in Ref.~\cite{pothier:96}. The self-consistent calculation presented
here leads to a slightly better fit than the curves derived in
Ref.~\cite{pothier:96}, in which a constant pair potential was assumed.  In
particular, the low-energy behavior of the experimental curves is reproduced
better.

\subsubsection{Minigap in a finite normal metal}
\label{fig:sec:minigap}
Next we consider a normal layer with finite thickness $d_{\rm N}\simeq\xi_{\rm
  S}$ in contact with a semi-infinite superconductor occupying the half space
$x<0$.  This system displays what is called a `minigap' in the density of
states of the normal metal.  This gap has been first noted by McMillan
\cite{mcmillan:68} within a tunneling model which ignores the spatial
dependence of the pair amplitude. We consider here the opposite limit,
assuming perfect transparency of the interface but accounting for the spatial
dependence of the Green's functions.

\begin{figure}[ht]
  \begin{center}
    \includegraphics[clip,width=10cm]{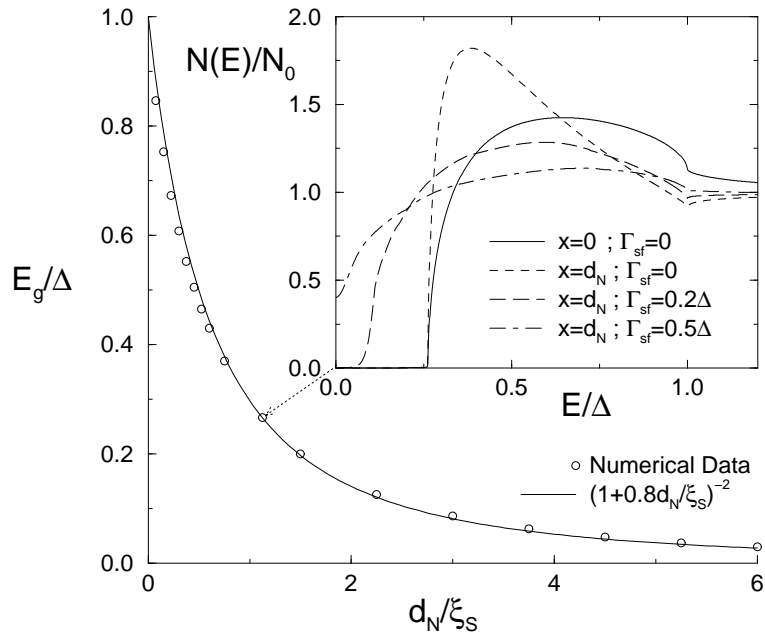}
    \caption[]{
      Minigap $E_{\rm g}$ as a function of the normal-layer thickness. Inset:
      local DOS of an N-layer of thickness $d_{\rm N}=1.1\xi_{\rm S}$ in
      proximity with an bulk superconductor .}
    \label{fig:minigap}
  \end{center}
\end{figure}

The boundary conditions at the interface to the superconductor remains
unchanged as compared to what has been discussed above. But at the open side,
$x=d_{\rm N}$, the condition is $d\theta(E,x)/dx = 0$, which is the same as if
the normal metal were bounded by an insulator.  In this case the DOS in the
normal layer develops a minigap at the Fermi energy. It is smaller than the
superconducting gap, and decreases as the thickness of the normal layer is
increased.  Results obtained from the self-consistent treatment are shown in
Fig.~\ref{fig:minigap}. Details of the shape of the DOS depend on the position
within in the N-layer \cite{belzig:96-2}. However, the magnitude of the
minigap is space-independent, as displayed by the inset of
Fig.~\ref{fig:minigap}.  The magnitude of the gap is expected to be related to
the Thouless energy $E_{\rm Th} \sim D/d_{\rm N}^2$, which becomes a relevant
energy scale in mesoscopic proximity system. This relation has to be modified
in the limit $d_{\rm N}\to 0$. Indeed, as shown in Fig.~\ref{fig:minigap}, a
relation of the form $E_{\rm g} \sim ({\rm const}\ \xi_{\rm S} +d_{\rm
  N})^{-2}$ fits numerically quite well. The sum of the lengths may be
interpreted as an effective thickness of the N-layer since the quasiparticle
states penetrate into the superconductor to distances of the order of
$\xi_{\rm S}$. The effect of spin-flip scattering in the normal metal on the
minigap structure is also shown in the inset of Fig.~\ref{fig:minigap}. The
minigap is suppressed as $\Gamma_{\rm {sf}}$ is increased until it vanishes at
$\Gamma_{\rm {sf}}\approx 0.4 \Delta$.  For $\Gamma_{\rm {sf}}=0$ our results
for the structure of the DOS agree with previous findings \cite{golubov:88,
  golubov:95}.  A minigap was also found in heterostructures with low
transparency barriers~\cite{volkovnato} and in a two-dimensional electron gas
in contact to a superconductor \cite{volkovetal94}. A minigap in the DOS has
also been found in a quantum dot coupled to a superconductor, if the shape of
the dot is chosen such that it shows chaotic behavior in the classical
limit~\cite{beenakker}. This systems are frequently described by random matrix
theory.  In contrast, dot structures which are integrable in the classical
limit have a non-vanishing DOS at low energy. This behavior is what one finds
in clean normal metal coupled to a superconductor \cite{stjames}.

\subsubsection{Overlap junction}
\begin{figure}[htbp]
  \begin{center}
    \begin{minipage}[b]{6cm}
      \includegraphics[clip,width=6cm]{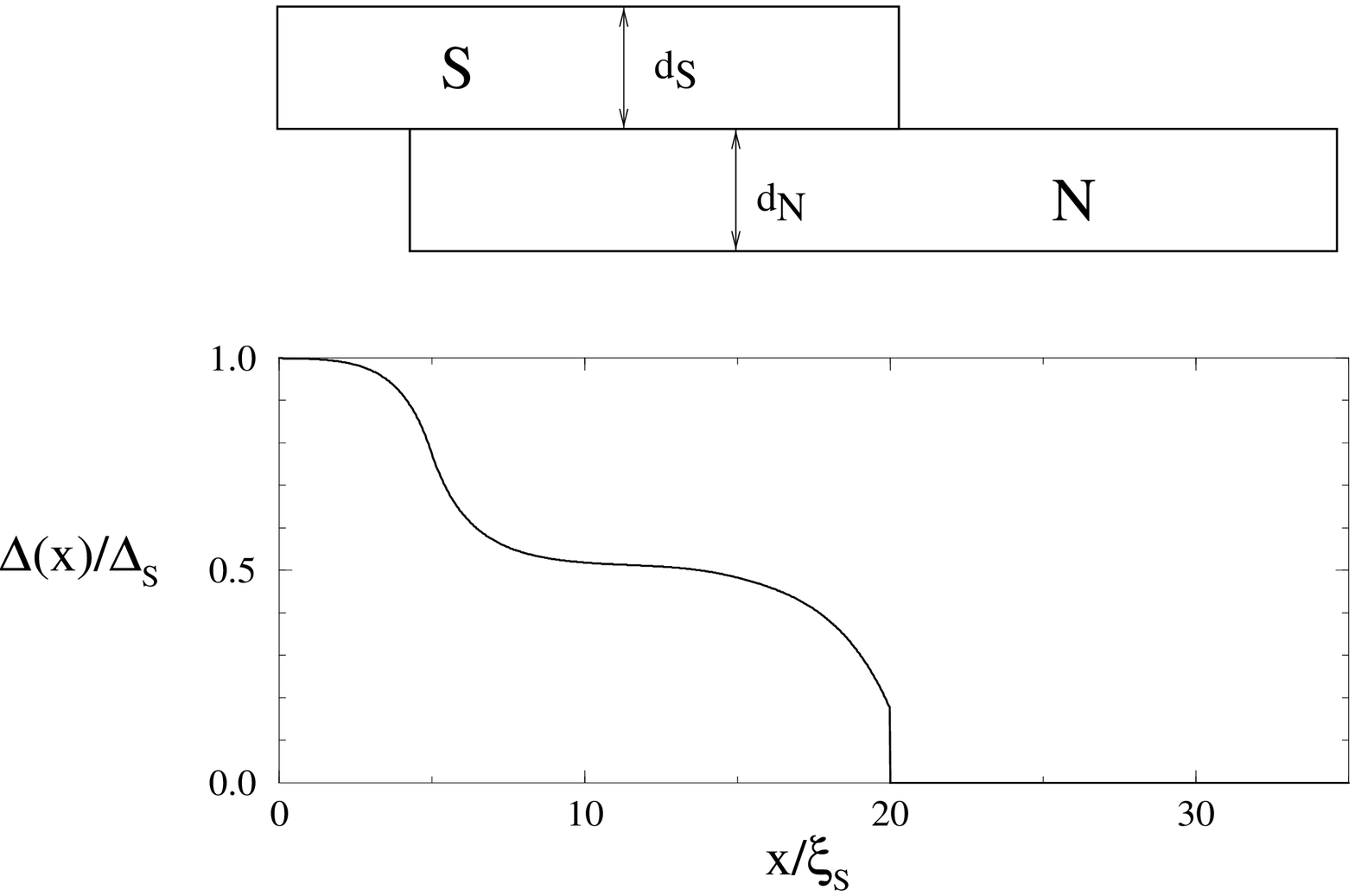}
      \caption{
        Geometry and pair potential of an overlap junction for $d_{\rm S}/(d_{\rm
          S}+d_{\rm N})=0.7$. The pair potential saturates around $0.5\Delta_{\rm
          S}$ in the middle of the junction.}
      \label{fig:overlap_delta}
    \end{minipage}
    \hfill
    \begin{minipage}[b]{8cm}
      \includegraphics[clip,width=8cm]{overlap.eps}
    \end{minipage}\\
    \caption[]{
      Density of states along an overlap junction. The curves for different
      positions are offset by $0.1$ on the vertical axis. The distance between
      two adjacent LDOS is $2.5\xi_S$ and the position is indicated by the bar
      on the left hand side.  One clearly sees the BCS-like density of states
      with reduced gap in the middle of the overlap junction.}
    \label{fig:overlap_dos}
\end{center}
\end{figure}

Finally we consider a geometry, in which the normal metal and the superconductor
have a finite overlap region in $x$-direction (see
Fig.~\ref{fig:overlap_delta}) which is longer than $\xi_{\rm S}$. This
geometry tries to mimic the experimental realization of junctions between
different metals. A measurement of the density of states along such a
junction would give additional information on the details of the proximity
effect. In principle the problem is two-dimensional. But if the total
thickness $d_S+d_N$ of the two layers is less than the coherence length, we can neglect
all derivatives in the transverse direction and simply average the Usadel
equation over the width.  This result for the self-consistently determined
pair potential is shown in the lower graph of Fig.~\ref{fig:overlap_delta}. In
the middle of the junction the pair potential saturates at $0.5\Delta_{\rm
  S}$.

The density of states along the overlap junction develops rather interesting
structures as shown in Fig.~\ref{fig:overlap_dos}. The DOS evolves gradually
from the BCS-behavior in the lowest curve to the normal metal behavior in the
topmost curve. In the middle of the junction a BCS-like peak in the density of
states appears at an intermediate energy due to the aforementioned 
plateau in the pair potential.\\

\subsection{Magnetic response of a proximity structure}
\label{sec:diamag}
The induced superconductivity enables a normal metal to produce magnetic
screening \cite{orsay:66,oda:80} reminiscent of the usual Meissner effect in
superconductors. However, it depends on mesoscopic scales, like the thermal
coherence length $\xi_T= v_{\rm F}/ 2\pi T$, and the thickness of the normal
metal $d_{\rm N}$
\cite{zaikin:82,mota:82,pobell:87,mota:89,mota:90,mota:94,higashitani:95,belzig:96-1,fauchere:97,belzig:98-3,belzig:98-2}.
In addition, a length related to the penetration depth of superconductors, the
London lenght$\lambda_{\rm N}^{2} \equiv 4\pi e^{2}n_{\rm e}/m$ plays an
important role.

In the following we we develop a linear response relation
between the current and the vector potential in terms of the zero-field
Green's functions in the presence of arbitrary impurity concentration. We will
show, that this formula reproduces previous results for the magnetic screening
obtained in the clean and dirty limit. Finally we show that for intermediate
strength impurity scattering qualitatively different results arise, consistent
with the observations in recent experiments.
 
The nonlinear response of the normal metal under proximity shows interesting
properties too \cite{orsay:66,belzig:96-1,fauchere:97}. At low temperatures,
the magnetic response exhibits a first order transition both in the clean and
in the dirty limit. At a certain critical field, called the breakdown field,
the induced magnetization is suddenly reduced, leading to a less diamagetic
state, or even vanishing diamagnetism in the clean case. Both, the breakdown
field and the size of the jump depend on temperature. All these features have
been measured experimentally \cite{mota:89}. The qualitative agreement of
the temperature dependence of the breakdown field with the clean limit theory
\cite{fauchere:97} showed that the samples are close to the ballistic regime.

\subsubsection{Linear response formalism}
We consider the magnetic screening in linear response of a 
system composed of a superconductor occupying the half-space $x < 0$,
in contact with a normal layer of finite thickness occupying the region
$0 \le x \le d_{\rm N}$, to a magnetic field 
$\bbox B (\bbox r)=B(x)\bbox{\hat z}$ in $z$-direction. We chose a gauge such
that the vector potential is $\bbox{A}(\bbox r)=A(x)\bbox{\hat y}$. 
For the analysis it is
convenient to expand the Green's functions in the Riccati parameterization
\begin{equation}
 \label{eq:riccati_expansion}
 a_\omega(v_x,v_y,x)=a^0_\omega(v_x,x)+ia^1_\omega(v_x,v_y,x)
\end{equation}
and a similar expansion for $a^\dagger$. Here $a^0$ and $a^1$ are the
first terms is in an expansion in $\bbox A$, and we have 
explicitly indicated the dependence on the two relevant components of
$\bbox v_{\rm F}$. Accordingly we expand the diagonal and off-diagonal
Green's functions $g_\omega$ and $f_\omega$. In first order the
Eilenberger equation reads
\begin{equation}
 \label{eq:riccati_linear}
  -\frac{v_{x}}{2}\frac{\partial}{\partial x}a^1_\omega(v_{x},v_{y},x)=
  \left[ \tilde{\omega}(x)+
   \tilde{\Delta}(x)a^0_\omega(v_{{x}},x)\right]
  a^1_\omega(v_{{x}},v_{{y}},x)+ 
  ev_{y}A(x) a^0_\omega(v_{{x}},x).
\end{equation}
The symmetry relation $a^1_\omega(v_x,v_y,x) = -a^1_\omega(v_x,-v_y,x)$, which
is a consequence of the gauge chosen, implies that the impurity self-energies
contained in $\tilde\omega$ and $\tilde\Delta$, see Eq.  (\ref{tildes}),
depend on $a^0$ only. Thus, we can integrate the linear inhomogeneous relation
(\ref{eq:riccati_linear}) directly.  The solution still depends on the
boundary conditions. Assuming perfect transparent N-S boundaries and
specular reflection at the boundary to the vacuum we obtain for the current
\begin{eqnarray}
 \label{eq:current_general}
 j_{y}(x)& =& -\frac{e^{2}p_{\rm F}^{2}}{\pi}T\sum_{\omega>0}
 \int\limits_{-\infty}^{d_{\rm N}}dx^{\prime}
 \int\limits_{0}^{v_{\rm F}}dv_x\frac{v_{\rm F}^{2}-v_x^{2}}{v_{\rm F}^{2}v_x}
 [1+g^0_\omega(v_x,x)][1-g^0_\omega(v_x,x^{\prime})]\\\nonumber&\times&
 \bigg[\Theta(x-x^{\prime})m_\omega(v_x,x,x^{\prime})+
 \Theta(x^{\prime}-x)m_\omega(-v_x,x,x^{\prime})
 + m_\omega(-v_x,x,d_{\rm N})m_\omega(v_x,d_{\rm N},x^{\prime}) \bigg]
 A(x^{\prime})\\ \nonumber
 &\equiv &-\int\limits_{-\infty}^{\rm d_N} K(x,x^\prime)
 A(x^{\prime})dx^{\prime}\; ,
\end{eqnarray}
where $g^0_\omega(v_x,x)$ is related to $a^0_\omega(v_x,x)$ by
(\ref{eq:riccati_parametrization}), and
\begin{equation}
 \label{eq:current_propagator}
  m_\omega(v_{x},x,x^{\prime})=\exp\left(\frac{2}{v_{x}}\int_{x}^{x^{\prime}}
   \frac{\tilde{\Delta}(x^{\prime\prime})}{
    f^{\dagger 0}_\omega(v_{x},x^{\prime\prime})}dx^{\prime\prime}\right)\; .
\end{equation}
The relation (\ref{eq:current_general}) is characterized by the interplay
between the range of the kernel determined by the decay of
$m_\omega(v_{x},x,x^{\prime})$ and the amplitude of the prefactor
$[1+g^0_\omega(x)][1-g^0_\omega(x^\prime)]$. This determines the non-locality
of the current-vector potential relation.

To examplify the usefulness of this result we reproduce the current response
of a half-infinite superconductor. Setting $d_{\rm N}=0$, the solution of the
Eilenberger equation (\ref{eq:eilenberger}) takes the simple form
$g_0=\omega/\Omega$, $f_0=f^{\dag}_0=\Delta/\Omega$, where
$\Omega=(\Delta^2+\omega^2)^{1/2}$. Inserting this in
(\ref{eq:current_general}) we obtain the linear-response kernel
\begin{eqnarray}
 \label{kernelsuper} 
 K_{S}(x,x^{\prime}) & =&\frac{e^{2}p_{\rm F}^{2}}{\pi}
 T\sum_{\omega>0}\frac{\Delta^{\!2}}{\Omega^2}
 \int\limits_{0}^{v_{\rm F}}\!\!du\frac{1\!-\!u^{2}/v_{\rm F}^{2}}{u}
 \big[
 e^{-(2\Omega+\frac{\scriptstyle 1}{\scriptstyle\tau}) 
 \frac{\scriptstyle|x-x^{\prime}|}{\scriptstyle u}}\!\!+
 e^{(2\Omega+\frac{\scriptstyle 1}{\scriptstyle\tau})
 \frac{\scriptstyle x+x^{\prime}}{\scriptstyle u}}\big]\; .
\end{eqnarray}
This result describes the current response of an arbitrary superconductor, as first
derived by Gorkov\cite{abrikosov:63}, which here additionally includes the
effect of the boundary. For fields varying rapidly spatially we arrive at a
non-local current-field relation of the Pippard-type\cite{pippard}, while for
slowly varying fields the kernel can be integrated out in
Eq.~(\ref{eq:current_general}), producing the local London result\cite{london}. 

The solution (\ref{eq:current_general}) is valid rather generally. We now
concentrate on a structure with a normal metal of thickness $d_{\rm N}$ in
contact with a semi-infinite superconductor. For simplicity we neglect the
penetration of the field into the superconductor by choosing $A(x=0)=0$. This
approximation leads to corrections to the induced magnetization of the normal
metal of order $\lambda_{\rm S}/d_{\rm N}$ only.  As the second boundary
condition we put $dA(x)/dx|_{x=d_{\rm N}}=H$ with the applied
magnetic field $H$. The spatial integration is now restricted to $[0,d_{\rm N}]$
and $m$ takes the form
\begin{equation}
 \label{eq:normal_propagator}
  m_\omega(v_{x},x,x^{\prime})= 
  \exp\left(\frac{1}{v_{x}\tau}\int_{x}^{x^{\prime}}
   \frac{\langle f^0_\omega(x^{\prime\prime})\rangle}{
    f^{\dagger 0}_\omega(v_{x},x^{\prime\prime})}
   dx^{\prime\prime}\right)\;.
\end{equation}
Here, $\langle f^0_\omega(x)\rangle=\int_{-1}^1duf^0_\omega(v_{\rm F}u,x)$ and 
$f^0_\omega(v_x,x)$ is related to $a^0_\omega(v_x,x)$ by
(\ref{eq:riccati_parametrization}).

The magnetic response of the normal-metal layer is expressed by the magnetic
susceptibility
\begin{equation}
  \label{eq:susceptibility}
  \chi=-\frac{1}{4\pi}\left[1- \frac{A(d_{\rm N})}{Hd_{\rm N}}\right]\;.
\end{equation}
It's calculation requires a self-consistent solution of
Eq.~(\ref{eq:current_general}) and the Maxwell equation,
$ d^{2}A(x)/dx^{2} =-4\pi j_{y}(x)$.
We will now present results for the clean and dirty
limits, as well as the intermediate regime.

\subsubsection{Clean limit}
In the clean limit, $\tau\to\infty$, and consequently $m=1$. Furthermore,
$g^0_\omega(v_x)$ is constant in space, and the current takes the form
\cite{zaikin:82}
\begin{equation}
 \label{eq:jclean}
 j_{\rm clean}=-\frac{1}{4\pi\lambda^{2}(T)\,d_{\rm N}}
 \int\limits_{0}^{d_{\rm N}}A(x)dx\; ,
\end{equation}
where
\begin{equation}
 \label{lambdaclean}
 \lambda^{2}(T)=
 \left\{
 \begin{array}[c]{ll}\displaystyle
 \lambda_{\rm N}^{2} &\; \mbox{for}\;
 T=0\\[5mm]\displaystyle
 \frac{\lambda_{\rm N}^{2} T}{12 \, T_{\rm
 A}}\exp\left(2\frac{T}{T_{\rm A}} \right) &\; 
   \mbox{for}\; T\gg T_{\rm A}
 \end{array}
\right.
\end{equation}
Here we defined $\lambda_{\rm N}^{2} \equiv 4\pi e^{2}n_{\rm e}/m$ and the
Andreev temperature $T_{\rm A}\equiv v_{\rm F}/(2\pi d_{\rm N})$.  The
extremely nonlocal form of the current-field relation (\ref{eq:jclean}) leads
to an overscreening for $\lambda(T)<d_{\rm N}$. This means, the magnetic
induction $B(x)$ changes its sign inside the normal metal, reaching for
$\lambda_{\rm N}\ll d_{\rm N}$ at the interface to the superconductor a value
as large as $-H/2$, which points in {\sl opposite} direction of the external
field. The temperature is given by the Andreev temperature $T_{\rm A}$, which
plays the same role as the Thouless energy in the diffusive case.

\begin{figure}[th]
 \begin{center}
   \includegraphics[clip,width=10cm]{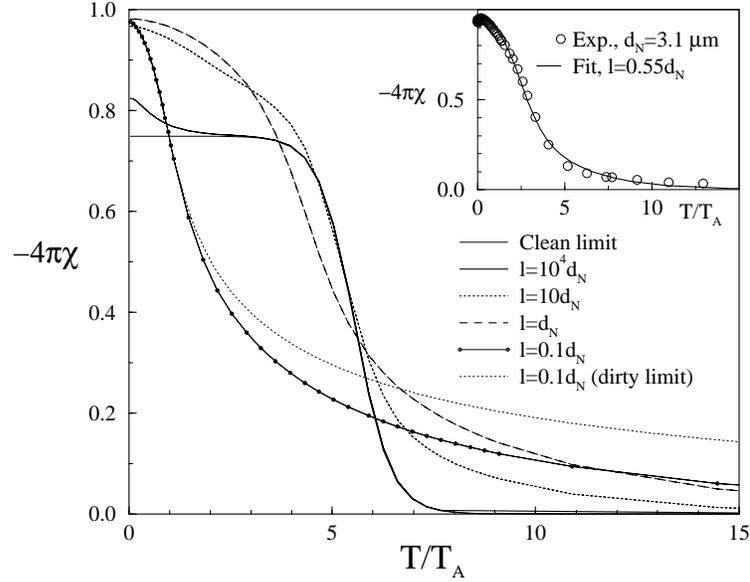}
   \caption[]{
     The susceptibility of the normal-metal layer for $\lambda_{\rm
       N}=0.003d_{\rm N}$.  The clean limit is indicated by a thin line
     reaching $0.75$ for $T\to 0$. Shorter mean free paths , $l$, (even as
     large as $10^4d_{\rm N}$) lead to an enhanced screening at low
     temperatures.  For shorter mean free paths, in the diffusive regime, a
     completely different temperature dependence is found. The dirty limit
     theory including a local current response deviates strongly from the
     correct result at higher temperatures. Inset: Comparison of experiment
     and theory.  The thickness of the normal metal in the experiment was
     $3.1\mu$m, corresponding to $T_{\rm A}=540$mK. The London length
     $\lambda_{\rm N}=22$nm was taken from the literature. To fit these
     samples a mean free path $l=0.55d_{\rm N}=1.7\mu$m had to be used.  This
     is in rough agreement with the result $l=4\mu$m of a transport
     measurement.}
   \label{fig:rho1} 
 \end{center}
\end{figure}

For the susceptibility we find
\begin{equation}
 \label{rhoclean}
 \chi=-\frac{1}{4\pi}\frac{3}{4+12\lambda^{2}(T)/d_{\rm N}^{2}}\; .
\end{equation}
In the limit $\lambda(T)\ll d_{\rm N}$ the susceptibility is $3/4$ of
$-1/4\pi$, thus the screening is not perfectly diamagnetic. For $\lambda(T)\gg
d_{\rm N}$ screening is exponentially suppressed. Both lengths coincide,
$\lambda(T)=d_{\rm N}$, at a crossover temperature $T \sim 2T_{\rm
  A}\log(d_{\rm N}/\lambda_{\rm N})$, which can be considerably larger than
$T_{\rm A}$ itself.  Interestingly, it is exactly this temperature below which
the first order transition in the nonlinear response occurs
\cite{fauchere:97}. The result for the linear susceptibility is displayed in
Fig.~\ref{fig:rho1} as the thin solid curve.

\subsubsection{Dirty limit}
Next we assume that the mean free path, $l$,  is the smallest length scale (except
for the Fermi wave length). In the absence of fields the Green's functions
vary on a scale $\xi_{\rm N}=(D/2\pi T)^{1/2}$, and the dirty limit is
realized if $l\ll \xi_{\rm N}$. However in the situation with a
space-dependent field, as seen from the kernel (\ref{eq:current_general}), a
stronger requirement is needed.  The dirty-limit form of the current-field
relation (\ref{eq:dirty_general_current}) is found only if vector potential
changes slowly on the scale of the mean free path.  Assuming this to be the
case we obtain the susceptibility shown in Fig.~\ref{fig:rho1} as the thin
dotted curve. The typical energy scale, below which the screening saturates is given
by the Thouless energy $E_{\rm Th}=D/d^2_{\rm N}$ in analogy to the Andreev
temperature $T_{\rm A}$ in the clean limit.

In order to check whether the requirement of a slowly varying vector potential
is satisfied, we define a local penetration depth $\lambda(x,T)$ as
\begin{equation}
 \label{eq:localpendepth}
 \frac{1}{\lambda^2(x,T)}=\frac{4\pi\tau}{\lambda_{\rm N}^2} 
 T\sum_{\omega>0} F_\omega^2(x).
\end{equation}
In order to evaluate it we can proceed with the approximate form
$F_\omega\sim\exp[-x(2\omega/D)^{1/2}]$.  As a result we find for the local
penetration depth
\begin{equation}
 \label{eq:usadelpendepth}
 \lambda(x,T)\approx
 \left\{
 \begin{array}[c]{ll}
 \lambda_{\rm N} x/l& 
 \mbox{if}\quad\xi_{\rm N}(T)\gg d_{\rm N}\\
 \lambda_{\rm N}\xi_{\rm N}(T) e^{x/\xi_{\rm N}(T)}/l&
 \mbox{if}\quad\xi_{\rm N}(T)\ll d_{\rm N}\;.
 \end{array}
 \right.
\end{equation}
For the local relation to be valid we need  $l<\lambda(x,T)$ in the
spatial region in which the screening occurs. For $T\ll E_{\rm Th}$ this means
$l\ll\lambda(d_{\rm N})$, leading to the requirement $l^2\ll \lambda_{\rm N}
d_{\rm N}$. The penetration depth $\lambda(d_{\rm N})$ is then given by
$\lambda_{\rm N} d_{\rm N}/l$. It is interesting to note, that this means that
full screening $\lambda(d_{\rm N})\ll d_{\rm N}$ is only achieved for $l\gg
\lambda_{\rm N}$. For $T\gg E_{\rm Th}$ screening takes place at
$x\approx\xi_{\rm N}$ and we have $l^2\ll\lambda_{\rm N}\xi_{\rm N}(T)$. The
low and high temperature conditions for the local approximation are different.
Consequently it is possible that the dirty limit theory can be applied  at
low temperatures, whereas it fails to describe the correct behavior at higher
temperatures.

A comparison of the dirty-limit result and the more general result for a
fairly short mean free path $l=0.1d_{\rm N}$, taking into account the nonlocal
screening, is shown in Fig.~\ref{fig:rho1}.  We clearly observe a difference
between the two results for higher temperatures. From the above discussion it
is clear that the temperature scale at which the deviation sets in is related to
the Thouless energy $E_{\rm Th}$, as demonstrated by the result in
Fig.~\ref{fig:rho1}.

\subsubsection{Arbitrary impurity concentration}
Finally we allow for arbitrary values of the mean free path. A qualitative
understanding may be gained from looking at the current vector potential
relation in the limit $l\gg d_{\rm N}$. In the limit $T\ll T_{\rm A}$ the
zeroth-order Green's functions are given by the clean limit expressions. We
approximate the kernel (\ref{eq:current_general}) by
\begin{equation}
 \label{eq:kernelapprox}
 K(x,x^\prime)=\frac{1}{8\pi \lambda^2(T)d_{\rm N}} 
 \left[e^{-|x-x^\prime|/l}+
 e^{-(2d-x-x^\prime)/l}\right]\; .
\end{equation}
Since $l\gg d_{\rm N}$, the exponentials may be expanded to first order. As a
result, we obtain two contributions to the current $j(x)=j_{\rm clean}+j_{\rm
  imp}(x)$.  The first contribution $j_{\rm clean}$ is given by the clean
limit expression Eq.~(\ref{eq:jclean}). The second is an impurity induced
contribution
\begin{eqnarray}
 \label{eq:jyimp}
 j_{\rm imp}(x)&=&
 \frac{-1}{8\pi \lambda^2(T)d_{\rm N}}\int_0^{d_{\rm N}}
 \frac{|x-x^\prime|+2d_{\rm N}-x-x^\prime}{l}A(x^\prime)dx^\prime\; .
\end{eqnarray}
This relations demonstrates when deviations from the clean limit become
important. It is clear that the impurities cannot be neglected, if
$j_{\rm imp}(x)$ is comparable to $j_{\rm clean}$. We estimate this by
calculating the two contributions to the current using the clean-limit vector
potential. Comparing the two contributions, we find that impurities can be
neglected, if
\begin{equation}
 \label{eq:supercleancondition}
 \lambda^3_{\rm eff}(T) \equiv \lambda^2(T)l \gg d_{\rm N}^3\; .
\end{equation}
This equation defines a new length scale, the effective penetration
depth $\lambda_{\rm eff}$. For the clean limit to be valid at
$T=0$ the condition $\lambda_{\rm eff}(0)>d_{\rm N}$ has to be
fulfilled, since in this case the screening takes place on the
geometrical scale $d_{\rm N}$. In the case $\lambda_{\rm eff}(0)\ll
d_{\rm N}$ the field is screened on a scale $\lambda_{\rm eff}$ and
the susceptibility is strongly enhanced in comparison to the clean
limit. Nevertheless, the clean-limit behavior may reappear at higher
temperatures, since $\lambda_{\rm eff}(T)$ grows with temperature.

For $T\gg T_{\rm A}$ the deviations from the clean limit can be calculated as
perturbation in the  impurity scattering. The
correction to $g^0_\omega$ leads to a finite superfluid density close to the
superconductor via the factor $1-g_\omega^0(x^\prime)\approx
(\xi_T/2l)\exp(-2x/\xi_T)$ in the kernel. The range of the kernel is
modified by the correction to $f^{\dagger 0}_\omega$, which is non-negligible
in this case. Since $f^{\dagger 0}_\omega\approx (\xi_T/2l)\exp(- x/\xi_T)$
we find for $T \gg T_{\rm A}$
\begin{equation}
 \label{eq:mcorr}
 m_\omega(v_{x},x,x^\prime)=\exp\left(\frac{1}{l}
 \int_x^{x^\prime}dx^{\prime\prime}
 \frac{f^0_\omega(x^{\prime\prime})}{
  f^{\dagger 0}_\omega(x^{\prime\prime})}\right) 
\approx\exp\left(2\frac{x^\prime-x}{\xi_T}\right)\; .
\end{equation}
In deriving this equation the $v_x$-dependence was neglected by putting
$v_x=v_{\rm F}$. The range of the kernel is now given by $\xi_T$, which
is strongly temperature-dependent. For the current we find
\begin{equation}
 \label{eq:currcorr}
 j(x) \approx \frac{1}{\lambda^3_{\rm eff}(0)} 
 \int_0^{d_{\rm N}} dx^\prime
 e^{-2(x^\prime+|x-x^\prime|)/\xi_T} A(x^\prime)\; ,
\end{equation}
again showing the importance of the length scale $\lambda_{\rm eff}$.  In
the limit $\lambda_{\rm eff}(0)\gg\xi_T$ the field cannot be screened on the
scale $\xi_T$, leading to a vanishing susceptibility.  However, if
$\lambda_{\rm eff}(0)\ll\xi_T$ the field can be screened on a length scale
smaller than $\xi_T$ and the susceptibility will be finite.

For arbitrary mean free paths we have solved the screening problem
numerically. The susceptibility as a function of temperature is shown
in Fig.~\ref{fig:rho1}. The different curves refer to the clean limit
and to mean free paths $l/d_{\rm N}=10^{4},10,1,0.1$. In this plot we
see the all the features discussed above. For large mean free paths
the screening is enhanced in comparison with the clean limit at low
and high temperatures, whereas the clean limit is restored in an
intermediate temperature range. For small mean free paths ($l=0.1
d_{\rm N}$ in the graph) screening is suppressed also in the
intermediate temperature range.

A comparison to experimental data from Ref. \cite{belzig:98-3} is shown in
the inset. Using the mean free path as the only fit parameter they
can be reproduced well by the theory. Furthermore
the fitted mean free path is in rough agreement with an independent
experimental estimate obtained from a resistance measurement of the same sample
above the critical temperature.\\

\subsection{Equilibrium supercurrent in a diffusive SNS junction}
\label{ch:sns}
The supercurrent through an SNS junction had been studied in the past mostly
on the level of the Ginzburg-Landau theory \cite{degennes:64}, which is valid
near $T_{\rm c}$ when the order parameter is small.  Recent experiments
entered the mesoscopic regime at low temperatures, where, e.g.\, the coherence
length $\xi_{\rm N}=\sqrt{D/2\pi T}$ is comparable to the sample
size $d$. Here a more general approach is needed.

\begin{figure}
 \centerline{\psfig{figure=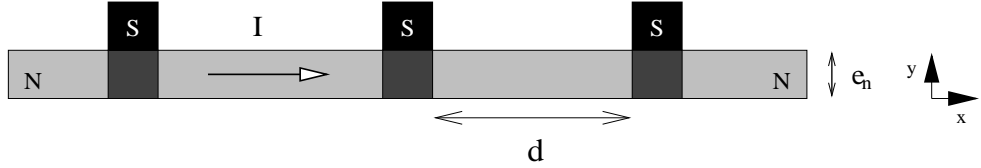,width=130mm}}
 \caption{
   A proximity wire as used in the experiment \cite{CourtoisPRB}. 
   The thickness is assumed to be $e_{\rm N}\ll\xi_0$ and $d$ is
   the distance between the strips}
\label{array}
\end{figure}

Strong deviations from the Ginzburg-Landau behavior have shown in the
current through a very thin (of thickness $e_{\rm N}$ much smaller than
$\xi_0$), diffusive normal wire on top of which an array of superconducting
strips had been deposited \cite{CourtoisPRB,CourtoisHere}, see Fig. \ref{array}. 
If there is a good
metallic contact between the superconductors and the wire, we can assume the
Green's functions under the strips to coincide \cite{KupriyanovSovjet} with
the values in the strips. These, in turn, can be assumed to be reservoirs, so
the Green's functions are given by their bulk values (\ref{Mat0}). The
structure is thus equivalent to a chain of SNS junctions without barriers
between the normal metal and the superconductor.

\begin{figure}
\begin{center}
\begin{minipage}{75mm}
\includegraphics[clip,width=75mm]{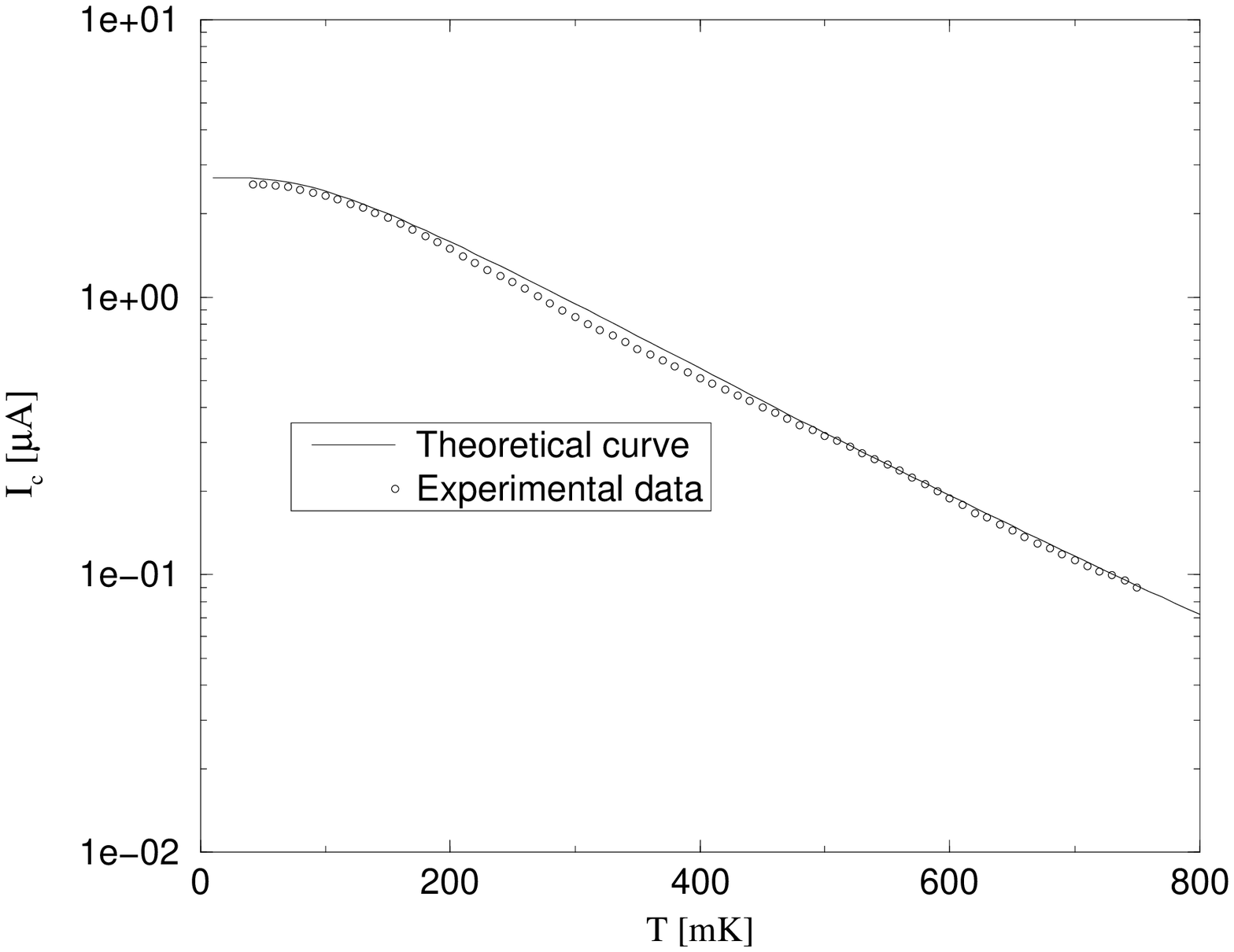}
\vspace{-7mm}
\caption{Critical current of the proximity wire compared with the experiment
\cite{CourtoisPRB}: $d=800$nm, resistance of the N-part between
two strips $R_{\rm N}=0.66\Omega$}.
\label{IC1}
\end{minipage}
\hfill
\begin{minipage}{70mm}
\vspace{-4mm}
\includegraphics[clip,width=68mm]{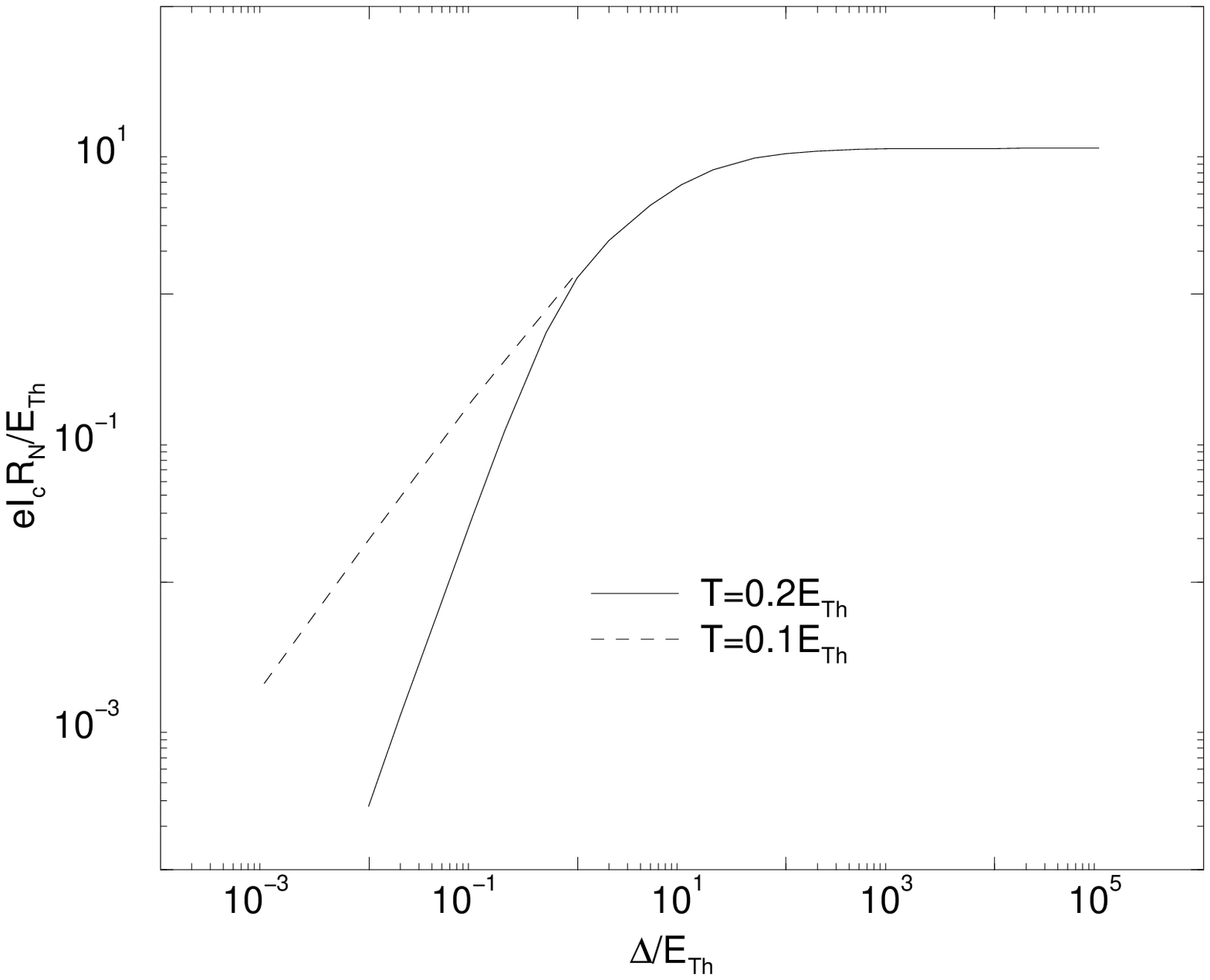}
\caption{The critical current at low temperature for different values of the
gap, normalized to the Thouless Energy. $\Delta\gg E_{\rm Th}$ corresponds to long and
$\Delta\ll E_{\rm Th}$ to short junctions.}
\label{IC2}
\end{minipage}
\end{center}
\end{figure}

We are interested in the equilibrium supercurrent. Hence, 
it is sufficient to use the Matsubara technique in the parameterization
(\ref{eq:theta_parametrization}). We thus have to solve the Usadel
equation  for the Green's function with the boundary
conditions (\ref{BCUsa}) assuming $\phi_0=\pm \phi/2$ at $x=\pm d/2$.  The
general structure of the solution is such that $\theta$ decays over a scale of
$\xi_\omega=\sqrt{{\cal D}/2\omega}$ from the boundaries. If $\xi_\omega\ll
d$, the overlap between the induced $\theta$ from the two superconductors is
exponentially small. Then we can approximate the pair amplitude 
by a superposition of the solutions describing a semi-infinite normal
part in the absence of phase gradients, i.e.
\begin{eqnarray}
F_\omega(x)&=&
e^{i\varphi/2}\sin\theta^+_\omega (x)
+e^{-i\varphi/2}\sin\theta^-_\omega(x) \; ,
\nonumber\\
\tan(\theta_\omega^\pm/4)&=&\exp\left(\pm x\sqrt{2\omega\over D}
+A_\omega^\pm\right)\label{HighE}\; .
\end{eqnarray}
For $T\gg E_{\rm Th}=D/d^2$ this approximate solution is sufficient to
calculate the current. From the sum (\ref{eq:matsubara_current}) we
find $I=I_{\rm c}\sin\phi$, where the critical current is
\begin{equation}
I_{\rm c}={64\pi T\over
eR_{\rm N}}\sum_{\omega}{d\over\xi_{\omega}}{\Delta^2\over{[\omega+\Omega
+\sqrt{2(\Omega^2+\omega\Omega)}]^2}}
\exp\left(-{d\over\xi_{\omega}}\right).
\end{equation}
At high temperatures the
sum over Matsubara frequencies is dominated by the first term. 
If, furthermore, $T\ll\Delta$, we arrive at
\cite{JLTP,ZaikinReview,ZaikinZharkov}
\begin{equation}
 I_{\rm c}={64\pi\over{3+2\sqrt{2}}}{T\over eR_{\rm N}}{d\over\xi_{\rm N}}
 \exp\left({-{d\over\xi_{\rm N}}}\right).
 \label{simpstrom}
\end{equation}
The exponent reflects the fact, that the weakest part of the structure, i.e.
the overlap region in the middle, defines the bottleneck for the supercurrent.

As becomes apparent from the semi-logarithmic plot in
Fig.~\ref{IC1} the critical current is approximately described by the 
numerical fit, $I_{\rm c}\propto e^{-T/T^\ast}$, with $T^\ast=12E_{\rm
Th}/\pi$.  This result arises because the convex exponential factor in
(\ref{simpstrom}), typical for diffusive metals, competes with the
concave pre-exponent $T^{3/2}$.  This surprising result resolves a
puzzle with the interpretation of the experiments of
Ref. \cite{CourtoisPRB}.  The observed linear $T$-dependence of the
exponent is what one would expect in a ballistic system
\cite{degennes:64,ZaikinReview}, whereas the experiments were
performed with diffusive metals.  The present result agrees well with
the experimental observation \cite{CourtoisPRB}.

At lower temperatures, all Matsubara frequencies contribute to
(\ref{eq:matsubara_current}) and we have to solve the Usadel equations
numerically. The results are shown in Fig. \ref{IC1}.  
As $T<E_{\rm Th}$, the critical
current saturates at a finite value $I_{\rm c}(T=0)$.  The
numerical results at very low $T$ (Fig. \ref{IC2} ) indicate that for a long
junction, $d\gg \xi_0$, the spatial decay of $F$  limits the
current to $I_{\rm c}R_{\rm N}\propto E_{\rm Th}$. In contrast, for a
short junction the gap is the relevant cutoff and $I_{\rm c}R_{\rm
  N}\propto\Delta$ \cite{Dubos}. This agrees qualitatively with
estimates derived from the low-energy expansion of the Usadel equation
\cite{JLTP} The numerical results for the critical current under
experimental conditions \cite{CourtoisPRB} explain the experiment
quantitatively well (see Fig.  \ref{IC1}).

\section{Applications to nonequilibrium transport problems}

\subsection{Supercurrent under nonequilibrium conditions: The SNS 
Transistor}
\begin{figure}[htb]
  \begin{center}
    \begin{minipage}[b]{5cm}
      \psfig{figure=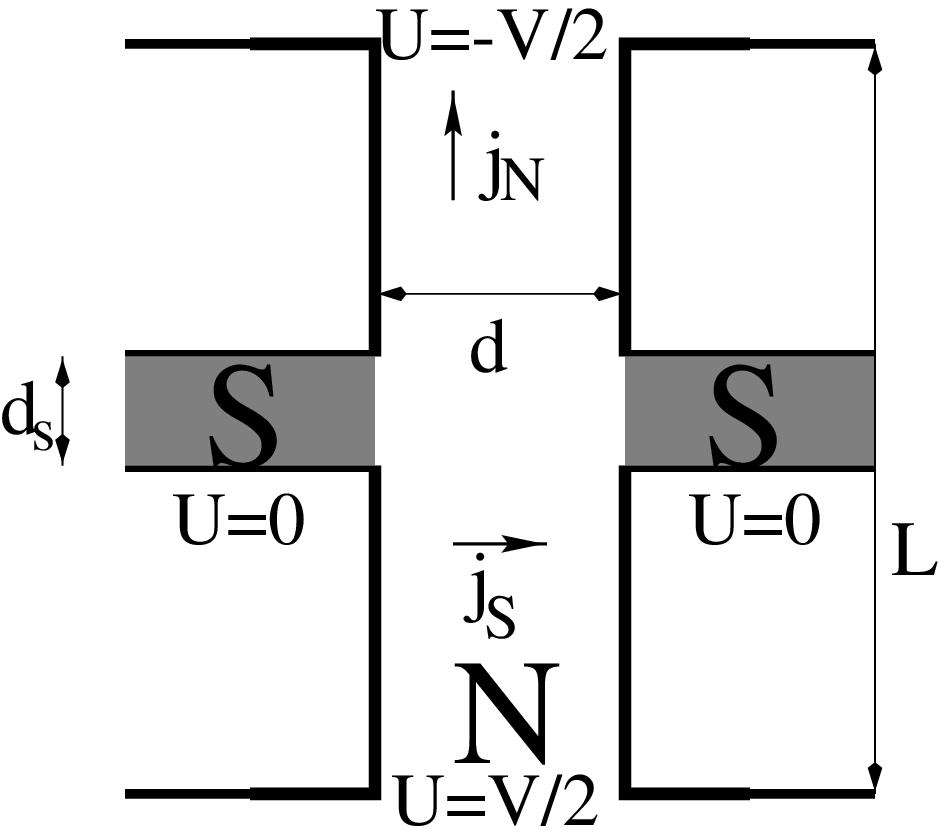,width=50mm}
    \end{minipage}
    \hspace{1cm}
    \begin{minipage}[b]{5cm}
      \psfig{figure=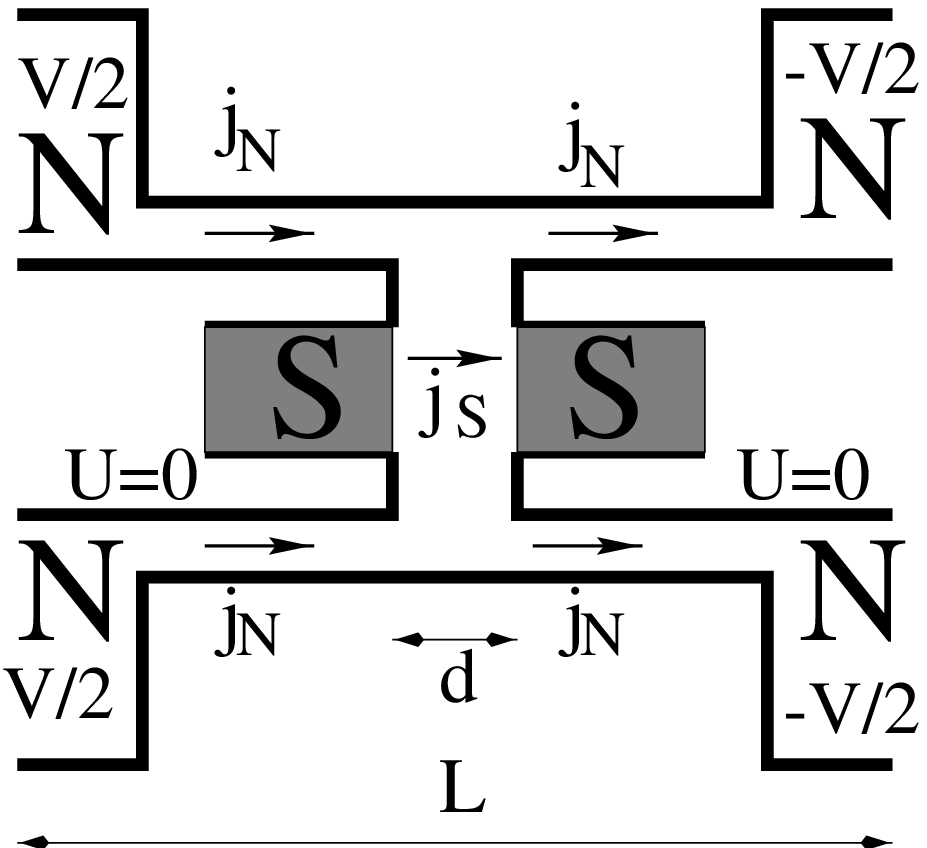,width=50mm}
    \end{minipage}
  \end{center}
  \caption[]{Different realizations of the SNS transistors. The 
    supercurrent is tuned by
    (a) a perpendicular or (b) a parallel normal current flow.}
  \label{transistor}
\end{figure}

In this section we will describe nonequilibrium transport problems. For this
purpose the Keldysh formalism outlined before is particularly appropriate. As
a first example we investigate how the supercurrent through an SNS structure
is influenced by a nonequilibrium current in the normal  part
\cite{WSZ}. Specifically we study the set-up which recently has been
investigated experimentally by the Groningen group
\cite{Morpurgo,Baselmans,WeesHere}. The device is shown in
Fig.~\ref{transistor}. Its center part is a normal  conductor connecting
massive reservoirs at voltages $\pm V/2$. The distribution function in this
part is obtained as a solution of the kinetic equation, which for a diffusive,
mesoscopic normal metal reads
\begin{equation}
    {\cal D}\partial^2_yf=0 \; .
\end{equation}
Its solution 
\begin{equation}
f(E,y) = \left(\frac{1}{2}-\frac{y}{L}\right)f^{\rm eq}
\left(E +eV/2\right)
+\left(\frac{1}{2}+\frac{y}{L}\right)f^{\rm eq}\left(E-eV/2\right)
\label{flres}
\end{equation}
has two temperature-rounded steps at the electrochemical potentials of both
reservoirs.  The step heights depend on the position along the wire; in this
way the distribution function interpolates linearly between the boundary
conditions at $y=\pm L/2$.  This functional dependence had been verified in
the experiments of Pothier {\sl et al.}~\cite{Pothier}.  At the center
of the
normal conductor superconducting electrodes are attached, through which a
supercurrent is flowing. By choosing a symmetric geometry we can assure that
the effective chemical potential of the distribution function at the site of
the superconducting contacts is zero. This guarantees that no net current is
flowing between the normal reservoirs and the superconducting leads. On the
other hand, the nonequilibrium distribution function influences the properties
of the SNS contact, and can be used to tune the supercurrent,
which creates the possibility to use the device as a transistor 
\cite{Morpurgo}. At low temperatures even a sign reversal of the
critical current, i.e. a transition to a so-called
$\pi$-junction has been
predicted \cite{WSZ}  and observed \cite{Baselmans,WeesHere}.

\begin{figure}[htb]
\centering
\begin{minipage}{72mm}
\psfig{figure=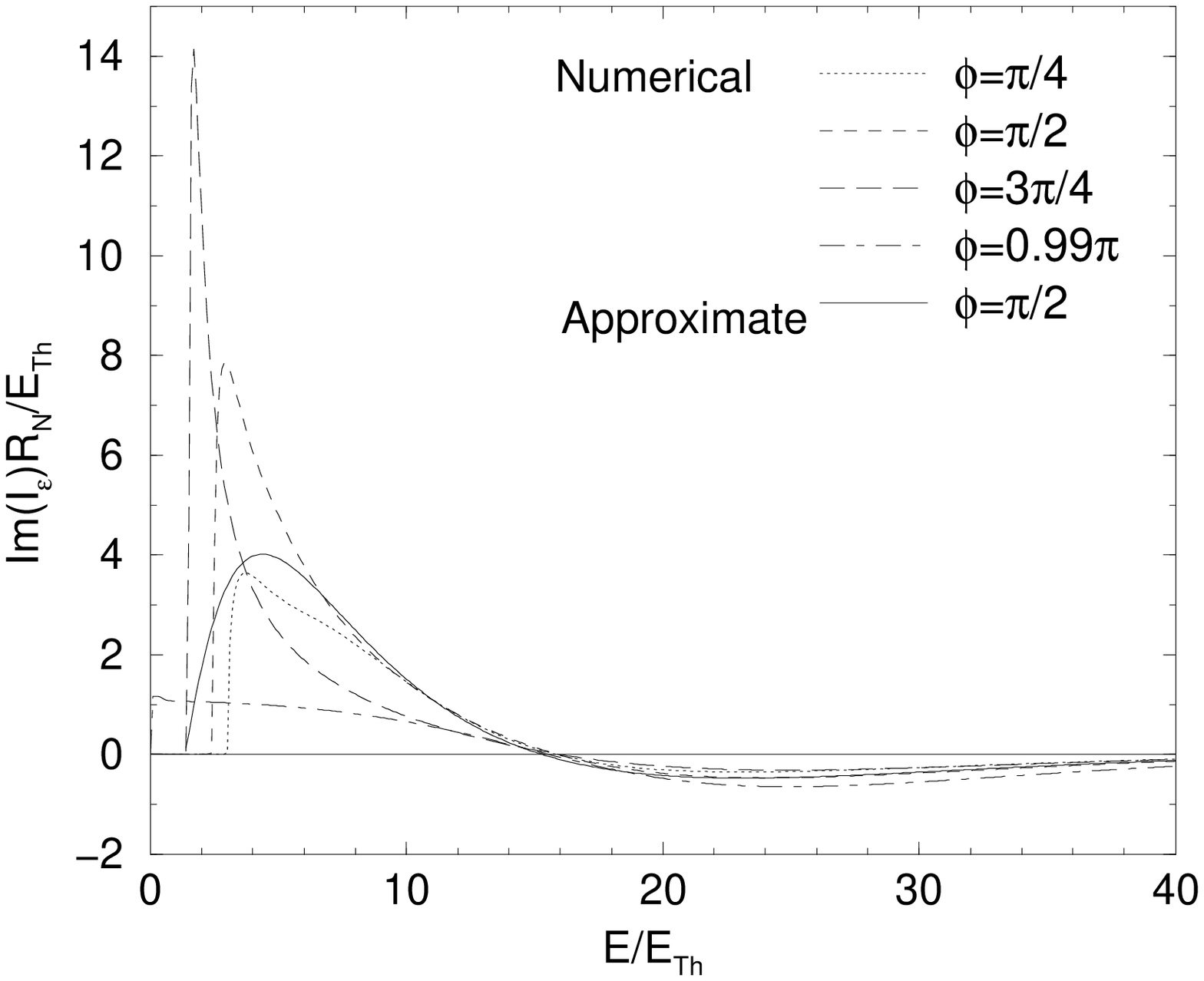,width=70mm}
\caption{The spectral current $\hbox{Im}(j_E)$ as a function 
of energy for different values of the phase difference $\phi$. At higher
energies we find oscillations as expected from eq. (\ref{TransAna}).}
\label{js1}
\end{minipage}
\hfill
\begin{minipage}{72mm}
\vspace{-3mm}
\psfig{figure=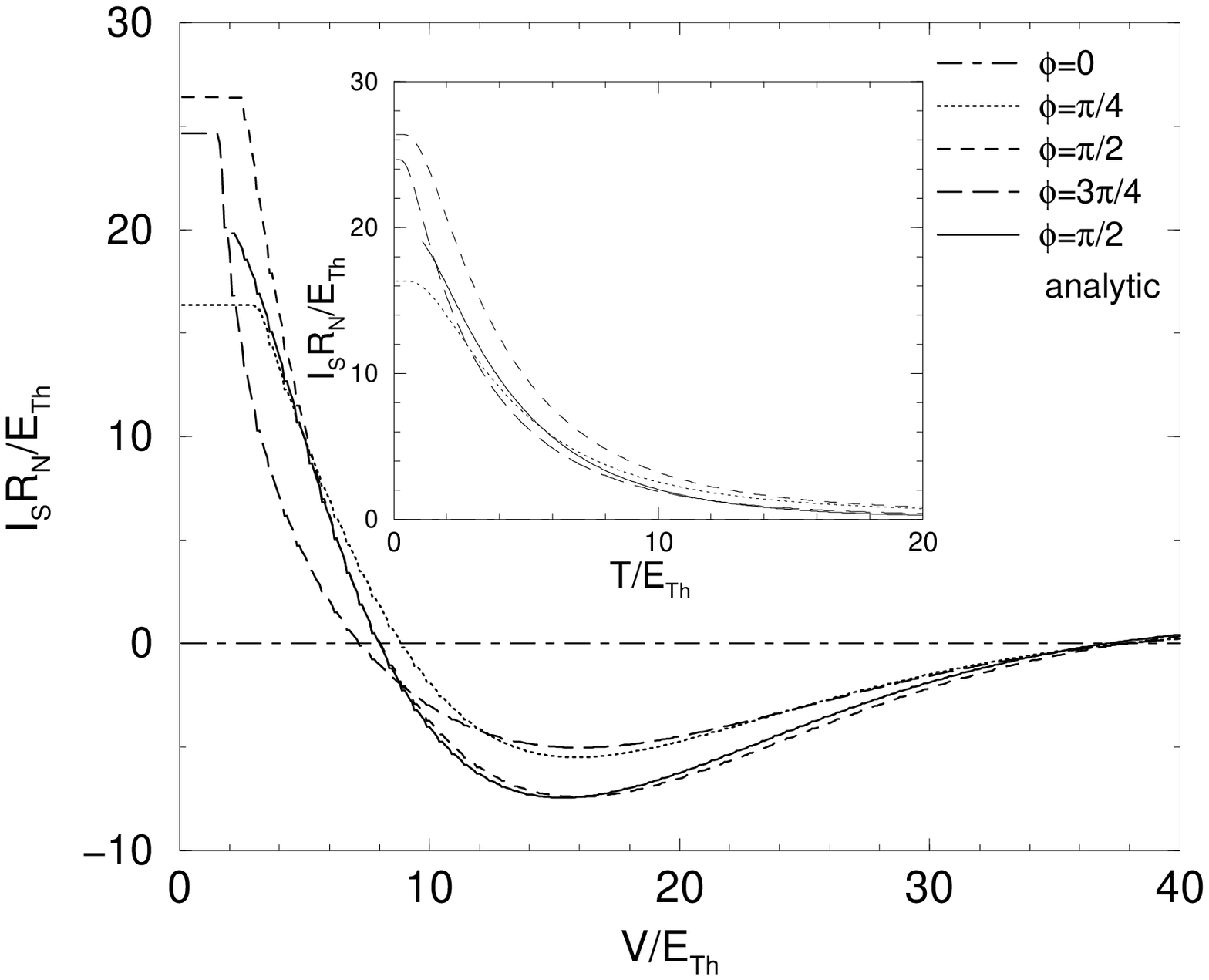,width=70mm}
 \caption{The supercurrent as function of control
voltage (at $T=0$) and temperature
(at $V=0$) for various values of $\phi$.}
\label{js2}
\end{minipage}
\end{figure}

To describe the system we have to evaluate the spectral functions 
as well as the distribution function. The former can be obtained from
Eq.~(\ref{retard}).  The low-energy ($E\ll E_{\rm Th}$) solution,
calculated perturbatively, reads
\begin{eqnarray}
 \theta &\simeq &-i\pi/2+(E/E_{\rm Th})\, a(\phi)+O(E^3)\nonumber \\
\chi &\simeq &\phi\, x/d-(E/E_{\rm Th})^2\,b(\phi)+O(E^3) \; ,
\end{eqnarray}
with real-valued functions $a$ and $b$.
This indicates that the spectral supercurrent $\hbox{Im} \{j_E\}$,
 vanishes in this energy range. At higher energies the problem 
can be analyzed numerically. However, more insight into the
problem is obtained from an approximate analytic solution. 
For the calculation of the spectral functions  between the
superconducting contacts we neglect the widening
of the geometry in the normal metal.  In 
this case we can use the solution (\ref{HighE}) obtained in 
the previous section. The approximation introduces errors of order one
in numerical coefficients, equivalent to a change in an effective area.
The result for the spectral supercurrent 
$\hbox{Im}\{j_E\}$, shown in Fig.~\ref{js1}, displays 
 a proximity-induced mini-gap
\cite{belzig:96-2,GolKupr} of size $E_{\rm g} \simeq 3.2 E_{\rm Th}$ at $\phi
=0$. This gap decreases with
increasing $\phi$, when the induced pair-amplitudes from both sides
start to interfere destructively, and vanishes at $\phi=\pi$
\cite{Charlat}. At energies directly above the gap, $\hbox{Im}\{j_E\}$
increases sharply, but rapidly decreases at higher $E$. At large
energies, it changes sign and eventually oscillates around zero with
exponentially decaying amplitude.

To evaluate the physical current (\ref{KeldyshSupercurrent}) we
combine the spectral supercurrent and the quasiparticle distribution
function $f$ between the reservoirs at electrochemical potentials $\pm
eV/2$ derived above. Although no net current is flowing between the
normal metal and the superconducting electrodes, the coupling might
influence the energy dependence of the distribution. We will now show
that this is not the case: Compared to the distance between the
reservoirs $L$ the proximity effect is confined to a relatively narrow
region $d_{\rm S}$. For a quantitative analysis we 
have to distinguish between the two
components $f_{\rm L/T}$, governed by the kinetic equations
(\ref{kinetic1}) and (\ref{kinetic2}). For the supercurrent
(\ref{KeldyshSupercurrent}) we need to determine only $f_{\rm L}$.
Because of the symmetry of the setup, the component $f_{\rm T}$
vanishes at the site of the superconducting leads.

As the voltage is applied between the reservoirs, the gradients of
$f_{L/T}$ point in $y$-direction, i.e. perpendicular to $j_E$ which
points in $x$-direction. As a consequence, the scalar products $\nabla
f_{L/T}\cdot j_E$ vanish and (\ref{kinetic1}) and (\ref{kinetic2})
decouple. Moreover, $f_{\rm L}$ satisfies the {\em same} boundary
conditions (\ref{KineticBoundary}) at the two reservoirs at $\pm
V/2$. Altogether, (\ref{kinetic2}) implies that $f_{\rm L}$ has a
constant value given by (\ref{KineticBoundary}) throughout the
wire. It is easy to check, using $2f=1-f_{\rm L}-f_{\rm T}$ with
$f_T=0$ and (\ref{KineticBoundary}), that the total distribution
combines to the double-step form described before.

At first glance, it may look surprising that $f_{\rm L}$ has such a
strong nonequilibrium form in the normal metal next to the two
superconductors, which are assumed to be in equilibrium
\cite{volkov:here}. It turns out to be a consequence of the
 symmetry properties of Andreev reflection. As the SN-interfaces are
good metallic contacts, the boundary conditions
eq. (\ref{eq:kuprianov_conditions1}) guarantee, that at the interface
${\cal D}_{\rm L}$ coincides with the value in the superconductor.
Thus at $|E|<\Delta$ we find from eq.  (\ref{diffkoeff}) that ${\cal
D}_{\rm L}=0$. This implies, that the SN-interfaces decouple
$f_{\rm L}$ in the normal metal from its value in the
superconductor. A well known
physical consequence of this property is, that below the gap, when
quasiparticles are transmitted by the Andreev reflection,
no heat is transferred \cite{andreev:64}.  
 The operation of the SNS-transistor without
tunneling barriers is based on this very idea \cite{Morpurgo}.

Our treatment of the 2D system is  approximate. A more complete
solution would also take the bending of supercurrent lines into account, which
leads to an effective renormalization of the cross section area. This 
problem has been solved by Volkov \cite{volkov:here} for 
a related system with small $d_S$ and tunneling barriers at
the SN-interface. An analysis of his solutions confirms the aforementioned 
argument, that the effective area of the normal region between the
superconductors is larger than $d_Nd$. This renormalization is reduces
at higher temperature and vanishes at $T\gg E_{\rm Th}$.

The form  of the distribution function implies
that in the energy window $-V/2< E < V/2$ current-carrying-states are
depopulated and the  supercurrent (\ref{KeldyshSupercurrent})
is strongly suppressed. For $V<E_g$ this
has no effect because of the gap. At higher voltage the supercurrent
decays rapidly with increasing voltage (cf. Fig.~\ref{js2}). This is exactly
what has been observed in the experiments
\cite{Morpurgo}. Furthermore, at still larger voltage $eV \ge 10
E_{\rm Th}$ the dominant contribution to  the integral in
(\ref{KeldyshSupercurrent}) outside this energy window is {\em negative} 
(see Fig.~\ref{js1},\ref{js2}), 
so the total supercurrent changes sign. We thus find
a transition to a $\pi$-junction \cite{Bul}, controlled by
nonequilibrium effects. This prediction has been verified in recent
experiments \cite{Baselmans,WeesHere}. The effect is rather
pronounced, the critical current of the $\pi$-junction is
approximately 30\% of $I_{\rm c}$ at $T=eV=0$. 

\begin{figure}
 \centering
\begin{minipage}{72mm}
\psfig{figure=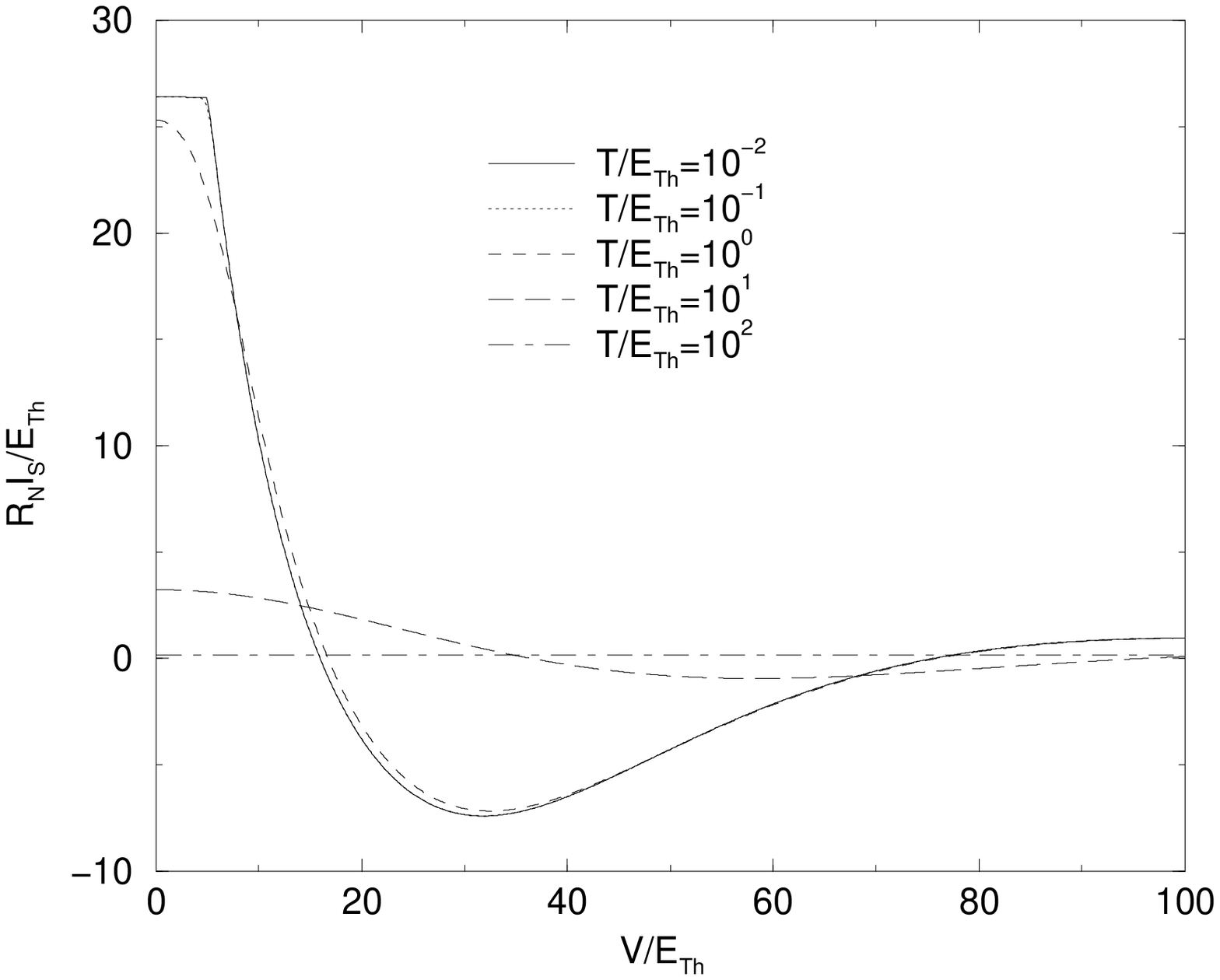,width=70mm}
\caption{The supercurrent as a function of the control
voltage for various values of $T$ at $\phi=\pi/2$.}
\label{moresup1}
\end{minipage}
\hfill
\begin{minipage}{72mm}
\psfig{figure=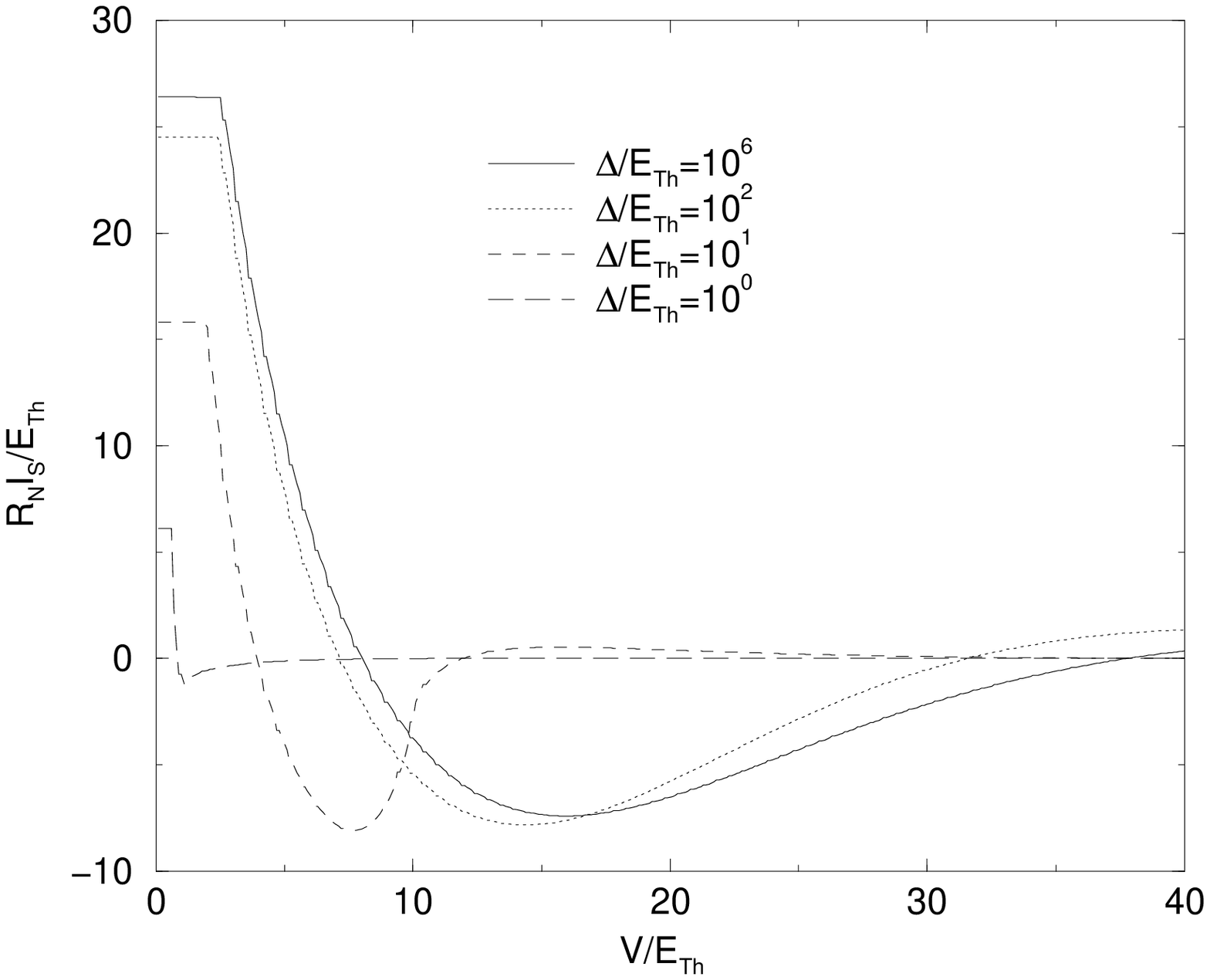,width=70mm}
\caption{The supercurrent as a function of the control
voltage (at $T=0$) for various values of $\Delta/E_{\rm Th}$
at $\phi=\pi/2$.}
\label{moresup2}
\end{minipage}
\end{figure}

For high voltages or temperatures $eV, T \gg E_{\rm g}$ we find a
closed expression for the current (\ref{KeldyshSupercurrent}). The integral is
dominated by the poles of $f_{\rm L}$, given by
(\ref{KineticBoundary}), at imaginary frequencies 
$E=i\omega_n\pm V/2$. Also $j_E$ can be found by 
analytic continuation of the approximation
(\ref{HighE}). We thus find 
$I_{\rm S}=I_{\rm c}\sin{\phi}$, with critical current
\begin{equation}
 I_{\rm c}R_d=\frac{64\pi}{3+2\sqrt{2}}
e^{-\sqrt{\Omega_V/E_{\rm Th}}}\cos\left(
\frac{eV}{2\sqrt{\Omega_VE_{\rm Th}}}+
\varphi_0\right)
\cases{eV&for $eV\gg E_{\rm g}$, $T\ll E_{\rm g}$\cr
T\left(\frac{{(2\pi T)^2+e^2V^2}}{E^2_{\rm Th}}\right)^{\frac14}&for 
$T\gg E_{\rm g}$\cr}\label{exp} \; .
\label{TransAna}
\end{equation}
Here $\Omega_V = \pi T+\sqrt{(\pi T)^2+(eV/2)^2}\label{exp2}$ and 
\begin{equation}
\varphi_0 = \cases{\pi/2&for $eV\gg E_{\rm g}$, $T\ll E_{\rm g}$\cr
\frac{1}{2}\tan^{-1}\left(\frac{eV}{2\pi T}\right)&for $T\gg 
E_{\rm g}$\cr}\; .
\end{equation}
This result indicates that with increasing temperature the onset of 
the $\pi$-junction behavior is weakened and shifted to higher voltages. 
Our numerical results, shown in Fig. \ref{moresup1},
demonstrate that it is still pronounced 
as long as $T\le E_{\rm Th}$.

Above we have discussed the limit of $\Delta\gg E_{\rm Th}$. 
For smaller values of $\Delta$  the minigap $E_g$ is limited by
$\Delta$ rather than $E_{\rm Th}$, see Fig. \ref{moresup2}. 
However, the qualitative behavior, including the transition to the
$\pi$-junction regime persists.\\

\subsection{Dissipative current in a mesoscopic wire}
\label{ch:reswire}
\begin{figure}
\centerline{\psfig{figure=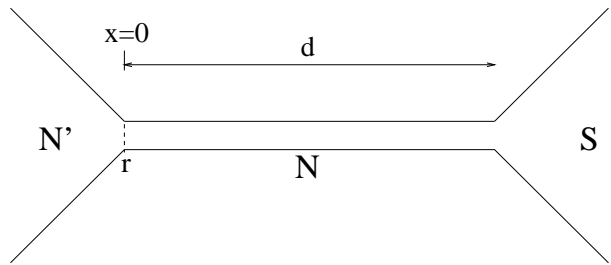,width=8cm}}
\caption{A thin normal wire between normal and superconducting reservoirs.}
\label{rw}
\end{figure}

Perhaps the simplest system displaying the influence of the proximity
effect on transport properties is shown in Fig.~\ref{rw}: A mesoscopic
wire connects a superconducting and a normal reservoir. At the
interfaces additional barriers (indicated by $r$ in Fig.~\ref{rw}) may
exist. In this section we assume no such barriers to be present.  We
assume the superconductor is assumed to be at voltage $V=0$.  As the
dissipative current is a genuine nonequilibrium quantity, we again
will use the Keldysh technique. In addition to the previously used
boundary condition at the superconductor, we impose $\theta=0$ at the
normal reservoir.  Hence, the spectral supercurrent $j_E$ vanishes at
the normal lead. As the retarded Usadel equation (\ref{retard})
conserves this spectral current, this implies that $j_E=0$ everywhere
and hence $\partial_x\chi=0$. This means that no `supercurrent' flows
in this system. On the other hand, the 'normal' current $\bbox{j}_n$
is influenced by superconducting correlations, and we have to study
their behavior in an electric field.

\begin{figure}
\centering
\begin{minipage}{72mm}
\psfig{figure=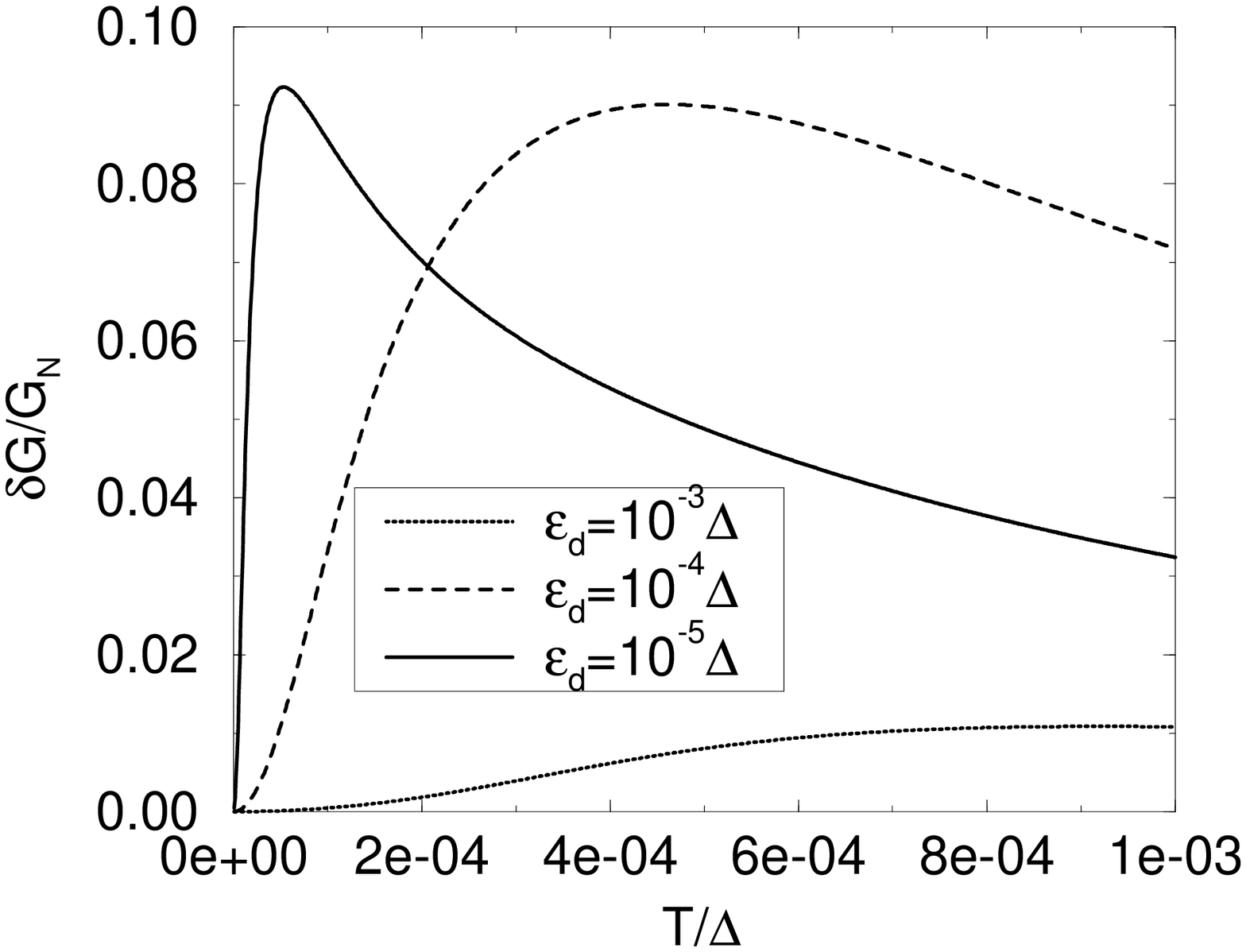,width=70mm}
\caption{Linear conductance of a proximity wire of length $d$ 
with transparent interfaces.The energy is
measured in units of the Thouless energy 
$E_{\rm Th}=D/d^2$ associated 
with the length $d$ of the wire.}
\label{dosfig1}
\end{minipage}
\hfill
\begin{minipage}{72mm}
\psfig{figure=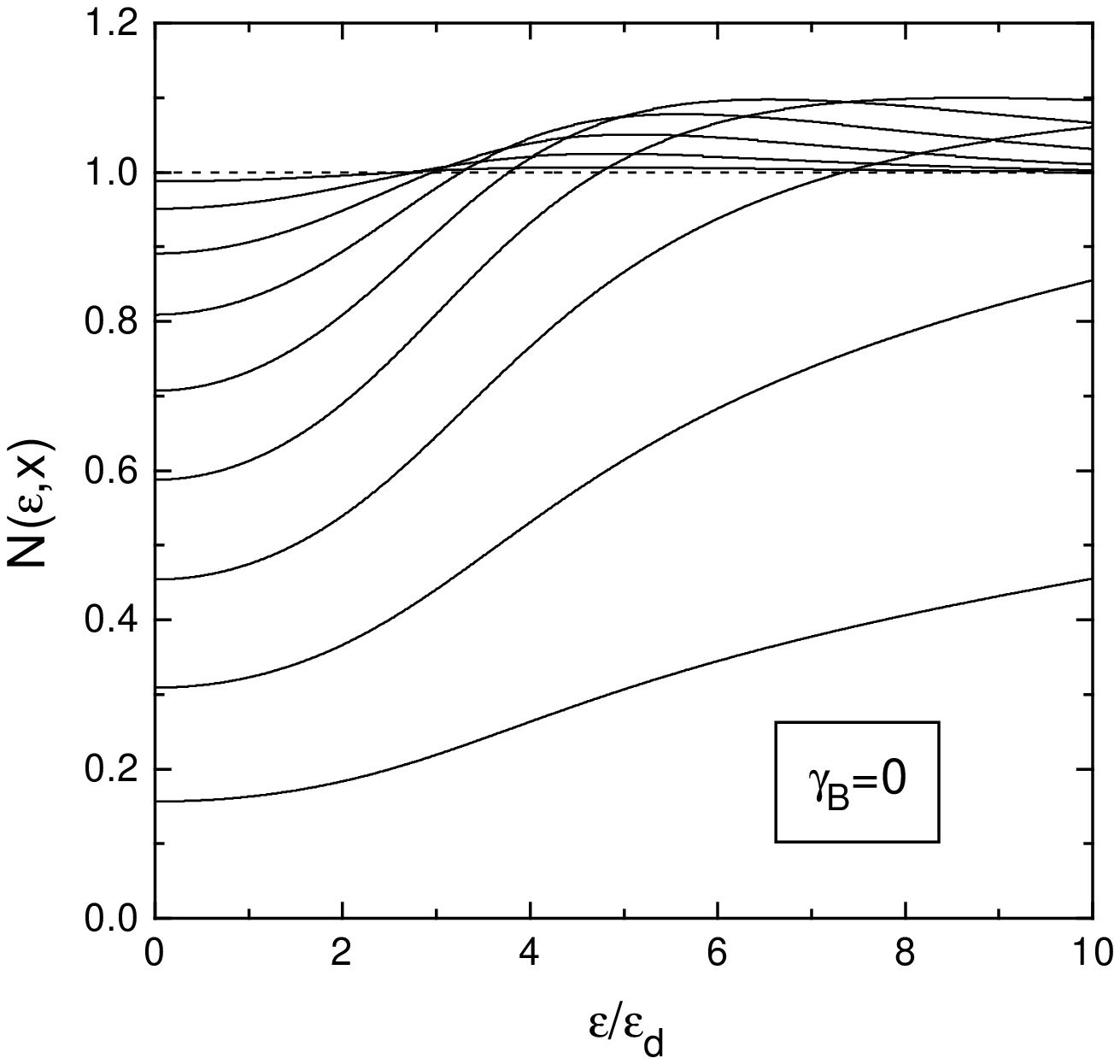,width=70mm}
\vspace{-3mm}
\caption{Local density of states in a proximity wire with transparent
interfaces for $\Delta \gg E_{\rm Th}$ at distances $x/d=0,0.1,\dots,1.0$
(top to bottom at low $E$) from the normal reservoir.}
\label{dosfig2}
\end{minipage}
\end{figure}

Approximate solutions of (\ref{retard}) can be obtained by the
same methods as described above. We will now concentrate on the case
$E_{\rm Th}\ll\Delta$. For $E\ll E_{\rm Th}$, iterating (\ref{retard}) from
the $E=0$-solution yields
\begin{equation}
\theta =-{\frac 8{\pi ^2}}{\frac E {E_{\rm Th}}}[\bar x-
\sin \left(\frac{\bar x\pi}{2}\right)]-i{\frac \pi 2}\bar x
\label{LowT}
\end{equation}
where $\bar{x}=x/d$. For $E\gg E_{\rm Th}$ we have
\begin{equation}
\label{HighT}
\tanh\left(\frac{\theta}{4}\right)=
-\tanh\left({\frac{i\pi}{8}}\right)\exp[k(x-d)] 
\end{equation}
with $k=\sqrt{-2iE/{\cal{D}}}$ .  For the calculation of the
dissipative current we only need the  distribution
function $f_{\rm T}$. As $j_E=0$, the kinetic equations decouple and
(\ref{kinetic1}) reduces to a simple spectral diffusion
equation. Relating $f_{\rm T}$ to the potential and ${\cal D}_{\rm T}$
to the conductivity, we note that our expressions (\ref{Dissip}) and
(\ref{kinetic1}) are analogous to Ohm's law and the continuity
equation on a spectral level, respectively.  It is now straightforward
to integrate (\ref{kinetic1}) and to calculate the current
(\ref{Dissip}), from which we obtain the differential conductance
\begin{equation}
G_=\int_{-\infty}^\infty \frac{dE}{2T\cosh^2 [(E-V)/2T]} 
g(E)\; .
\label{ZaitsevCond}
\end{equation}
Here the spectral conductance $g(E)$ is defined by
\begin{equation}
g(E)=G_{\rm N}d\left(\int_0^d\frac{dx}{{\cal D}_{\rm T}(E,x)}
\right)^{-1}\;.
\end{equation}
Again this expression
is the spectral form of Ohm's law. Using the approximate
solutions (\ref{LowT}) and (\ref{HighT}), we calculate ${\cal
D}_{\rm T}$ and $g(E)$, and arrive, e.g.\ at $V=0$, at
\begin{equation}
\frac{G}{G_{\rm N}}=1+\cases{a \, T^2/E _d^2 &for $T\ll E_{\rm Th}$\cr
b\, \sqrt{E_{\rm Th}/T}&for $T\gg E_{\rm Th}$\cr}
\end{equation}
with $a\approx0.049$ and $b\approx0.42$. Results are displayed in
Fig.~\ref{dosfig1}.  The conductance approaches the normal-state value for
$T\gg E_{\rm Th}$, because in this limit the proximity effect is confined to a
short portion of the wire of length $\xi_{\rm N}$. However, also at
$T\rightarrow0$ the correction to the conductance vanishes. This has been
first noted by Artemenko {\sl et al.} \cite{AVZ} for short contacts, and later
in longer wires by Nazarov and Stoof \cite{NazStPRL,NazStPRB,GWZ}. The theoretical
prediction was confirmed in experiments \cite{CourtoisHere,CourtoisLT}.  To
the best of our knowledge, no simple physical explanation of this surprising
coincidence has been provided yet. The fact that the conductance is unchanged
does {\em not} imply that the proximity effect is absent. Even at $E=0$
the solution (\ref{LowT}) yields $F^{\rm R}\ne 0$, and the density of states
$\hbox{Re}(G^{\rm R})$ displays a proximity-induced gap-structure, as shown in
Fig. \ref{dosfig2}.

At $E=0$, the local spectral diffusion coefficient ${\cal D}_{\rm T}$
coincides with the normal-state value everywhere, see
Eq.~(\ref{LowT}). Returning to the definition (\ref{DiffusionT}) and
using the parameterization (\ref{eq:theta_parametrization}) we find here
\cite{GWZ}
\begin{equation}
{\cal D}_{\rm T}=(\hbox{Re} G^{\rm R})^2+(\hbox{Im} F^{\rm R})^2=\cosh^2(\hbox{Re}(\theta))
\ge1\; .
\end{equation}
This means, that it is a combination of the density of states and the
Maki-Thompson paraconductivity term \cite{Tinkham,MakiThompson}. Both
terms compete. At finite $E$, the Maki-Thompson contribution is
stronger than the suppression of the DOS \cite{GWZ}. At $E=0$ both
contributions, although each of them varies in space, add up to 1.

In order to study the space-dependence more thoroughly, we look at the local
DOS in our model. Our results (\ref{dosfig2}) indicate, that due to the
induced superconductivity, a pseudo-gap in the DOS opens, i.e. the DOS is in
general suppressed but does not vanish at low energies.  If we move away from
the superconductor, the pseudo-gap is weakened and the DOS approaches its
normal-state form when we are close to the normal reservoir.\\

\subsection{Supercurrent and interference effects on the dissipative current}
\label{ch:oszi}
We will extend our study of the dissipative current to a situation
where also a supercurrent can flow and therefore interference of the
induced correlations starts to play a role. This type of effects was
first studied experimentally \cite{PetrPRL95} in cross-shaped
diffusive metal structures with different superconducting leads closed
by a SQUID loop, later also in analogous semiconducting structures
\cite{Hartog}, as well as in metallic structures with an Aharonov-Bohm loop
\cite{CourtoisPRL95,CourtoisPRL96}. Usually, these structures are
referred to as ``Andreev interferometers''. The common key observation
is, that the conductance depends periodically on the  magnetic
flux penetrating the loop with a period of a 
superconducting flux quantum $\phi_0=h/2e$. Moreover, it turned out
that these oscillations could be detected even if $\xi_{\rm N}\ll d$
where $d$ is a typical geometrical sample dimension. I.e., the
the proximity effect had a longer range than expected from 
simpler arguments. 

\begin{figure}
\centerline{\psfig{figure=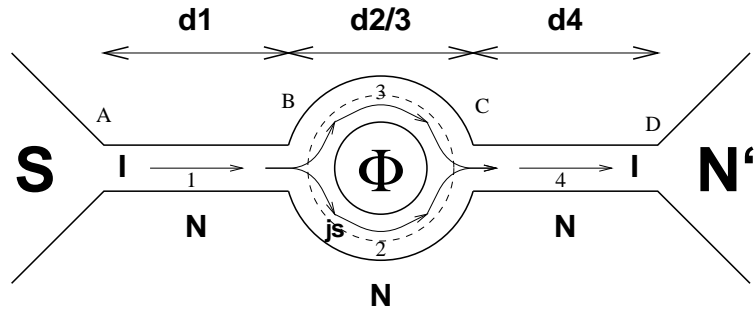,width=10cm}}
\caption[A proximity wire containing a loop]{A proximity wire containing a 
loop. The cross-section areas
of the respective components are $A_i$. {\boldmath $j^S$} is the
field-induced supercurrent in the ring part.}
\label{loop}
\end{figure}

These effects have been extensively studied theoretically
\cite{raimondi,NazStPRL,NazStPRB,GWZ,Stoof,Lambert}. We want to
outline the main concepts in these systems, considering as an example
the experiments of Ref. \cite{CourtoisPRL95}. The layout of
the system is shown in Fig.~\ref{loop}. 
Applying the arguments of the previous chapter, we recognize that the
supercurrent is confined to the loop. Instead of dealing with the
vector potential in the derivatives in (\ref{eq:usadel_isotropic}), we
remove it by a gauge transformation $\chi=\chi_0-2e\int_0^{\rm R}
dr^\prime\;\bbox{A}(r^\prime)$ where the integral runs around the
loop. This implies, that the vector potential produces an artificial
phase difference of $\phi=\Phi/\Phi_0$ (modulo $2\pi$) at one point in
the loop, which we choose to be point $B$. This procedure already guarantees
the $h/2e$-periodicity of the results.

We know from the previous sections the form of the
Usadel equations in the single branches. In the present problem we
also have to account for the nodes. To do so we apply the idea of Zaitsev
\cite{ZaiNet}:  By regarding the 
Green's functions in the connecting wires as cross-sectional averages of
the full 3D Green's function, which have well-defined derivatives in
space, we will conserve the ``spectral flow''. This means
\begin{equation}
\check{G}_n=\check{G}_m
\label{branch1}
\end{equation}
for all branches $n$ and $m$ and at nodal points we have
\begin{equation}
\sum_nA_n\check{G}{\partial\over\partial x_n}\check{G}=0\; ,
\label{branch2}
\end{equation}
where $n$ labels the branches, and the
derivatives point  in the direction of branch $n$ away from the
node. This approach has been used by Nazarov \cite{NazCirc,NazHere} to
derive a simple circuit theory for $T=V=0$. Here, we will assume that
the branches of the loop have {\em half} of the cross-section area of
the other branches.

Only very few analytical results have been derived for this system 
\cite{GWZ,Stoof}. Here we will focus on numerical results. 
\begin{figure}[htb]
\centering
\begin{minipage}{72mm}
\psfig{figure=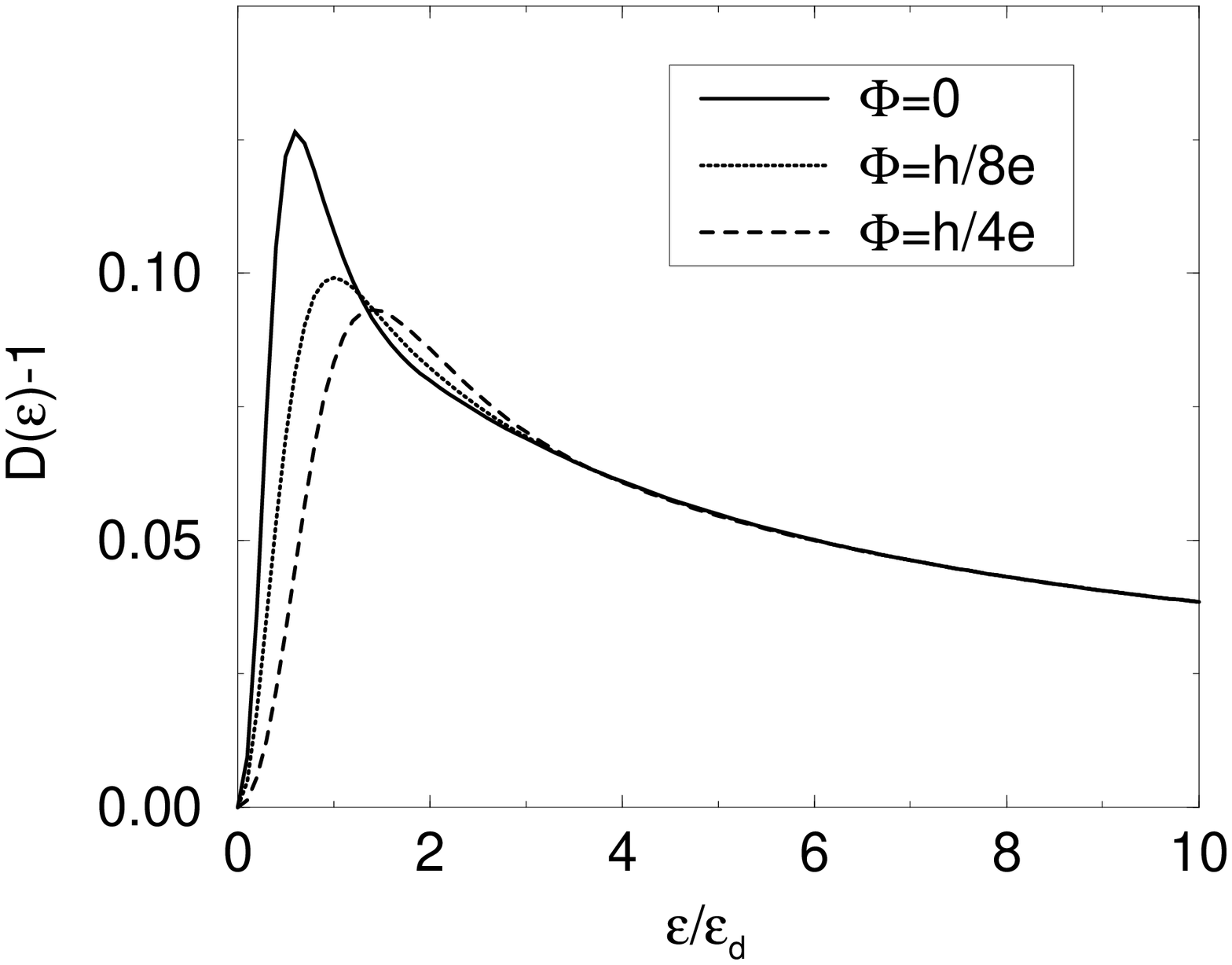,width=70mm}
\caption{Flux-dependent spectral conductance of the loop structure, assuming
$A_{1/4}=2A_{2/3}$ all $d_i=d$ and $\Delta\gg E_{\rm d}=D/d^2$.}
\label{spectralcond1}
\end{minipage}
\hfill
\begin{minipage}{72mm}
\vspace{-3mm}
\psfig{figure=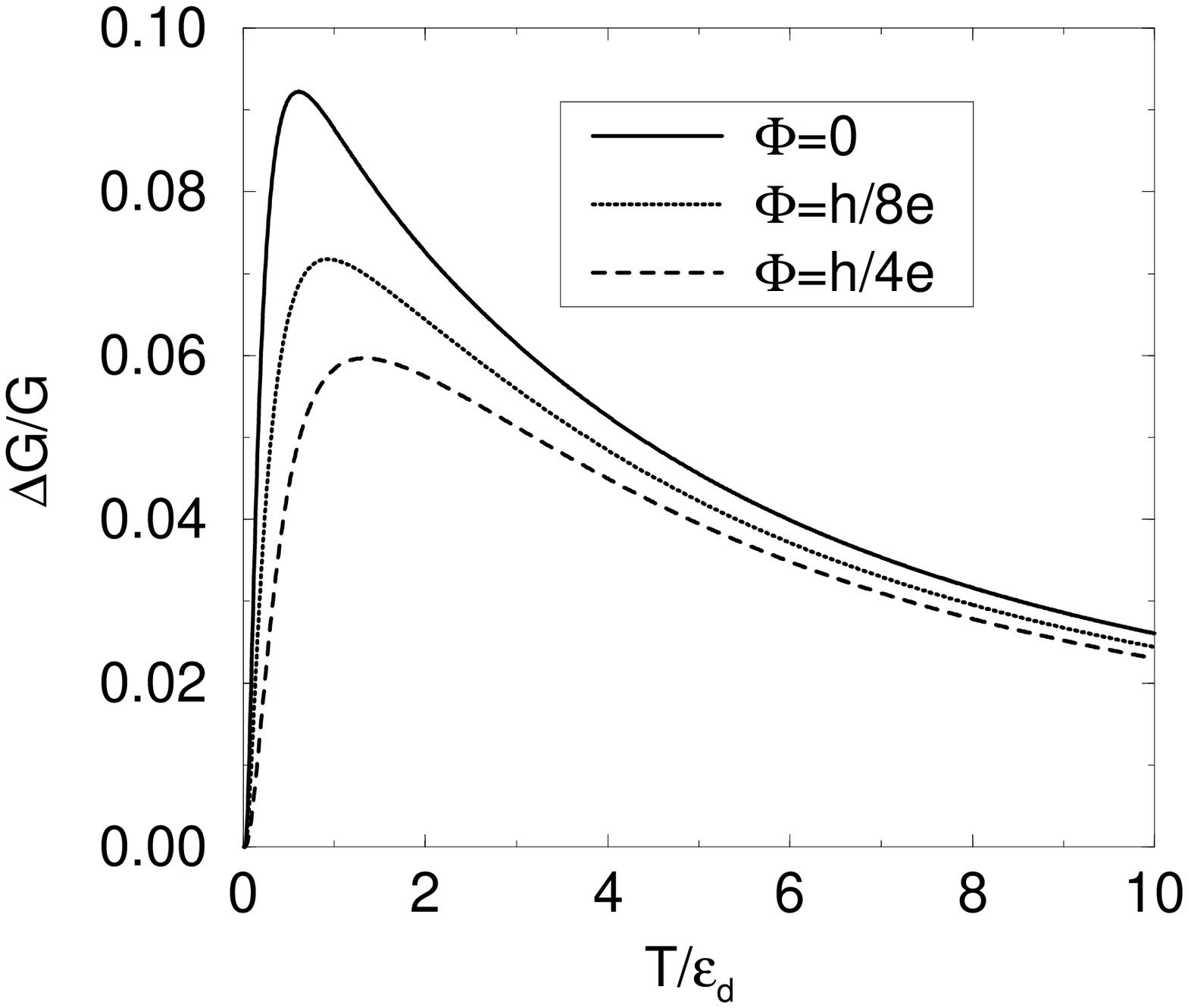,width=70mm}
\caption{Linear-response conductance of the loop as described in
the caption of Fig. \ref{spectralcond1}.}
\label{spectralcond2}
\end{minipage}
\end{figure}
As shown in Fig.~\ref{spectralcond1}, the spectral conductance shows a
non-monotonic shape as in the previous section.  Provided that the
induced correlations persist from the superconducting electrode to the
loop, the low energy results, for $E\ll E_{\rm Th}$, depend on the
flux, whereas the conductance at $E\gg E_{\rm Th}$ is
flux-independent, see Fig.  \ref{spectralcond1}. This agrees with the
intuition on the range of the proximity effect developed in the
previous sections. In the loop, the vector potential causes the
correlations to interfere destructively, so the induced conductance
correction is suppressed by the presence of the loop. At $E\gg E_{\rm Th}$, 
the correlations have already decayed before they reach the loop and are
therefore not influenced by the flux.
 On the other
hand the temperature dependent conductance also depends notably on the
flux at $T>E_{\rm Th}$, see Fig. \ref{spectralcond2}, so we indeed
recover the observed long range proximity effect
there\cite{CourtoisPRL95,PetrJETPL93,PetrPRL95}.  These seemingly
paradoxal observation can be resolved by examining (\ref{ZaitsevCond}).
In order to calculate the conductance Fig.~\ref{spectralcond2} from
the spectral conductance Fig.~\ref{spectralcond1}, we have to
convolute the latter with the derivative of the Fermi function.  This
distribution function is peaked around $E=0$ and has the width $T$,
i.e.  the window of size $E_{\rm Th}$ around the Fermi surface where
the spectral conductance is actually flux-dependent with a relative
weight of $E_{\rm Th}/T$.  Thus, the conductance depends sensitively
on the contribution of this low-energy range, where the proximity
effect has a long range~\cite{CourtoisPRL95}, which is only limited by
the phase-breaking length $l_\phi$.

\begin{figure}
\centerline{\psfig{figure=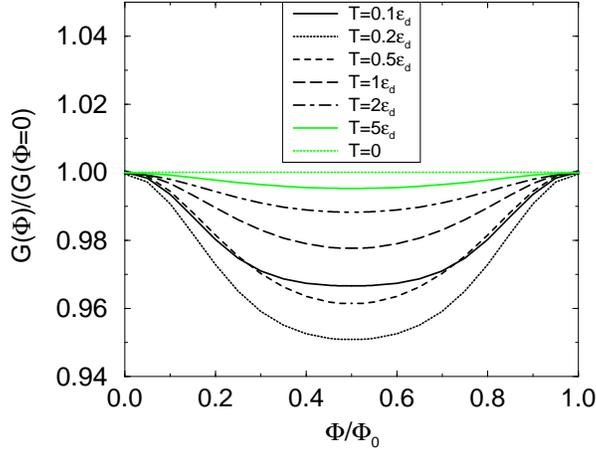,width=8cm}}
\vspace{-4mm}
\caption{Full period of normalized conductance oscillations in the loop 
structure at $V=0$.}
\label{fluxdep}
\end{figure}

We note that the intuition on $\xi_{\rm N}$ developed by regarding
thermodynamic quantities as the supercurrent in an SNS junction
(section \ref{ch:sns}) fails when describing nonequilibrium
quantities.  On the other hand, we have shown that the same
microscopic mechanism (characterized by the same Green's functions)
may lead to the exponential temperature dependence of equilibrium
quantities as well as to the ``long range'' tail with power-law
temperature dependence of nonequilibrium phenomena.

\subsection{Transport through tunneling barriers}

So far, we only considered situations, where the current flow does not
have to overcome tunneling barriers. We found, that the local spectral diffusion
coefficient ${\cal D}_{\rm T}=\cosh^2(\hbox{Re}\theta)\ge 1$ and
thereby the global conductance is always {\em enhanced} in comparison
to the normal state. However, as we explained 
before (section \ref{ch:DOS}), the tunneling conductance is proportional to
 the density of states, i.e. at low energies,
$E<E_{\rm Th}$, it is suppressed in comparison to the normal state as a
manifestation of the induced superconducting gap
\cite{GolKupr,belzig:96-2,pothier:96}. In general, these two mechanisms compete.
Moreover, due to the direct coupling to the
normal reservoir, correlations immediately feel the presence of normal
electrons and only a pseudo-gap develops, which turns into a real
minigap when the reservoir is decoupled from the system via a
tunneling barrier \cite{GWZ,ATWZ}.

\begin{figure}
\centering
\begin{minipage}{72mm}
\psfig{figure=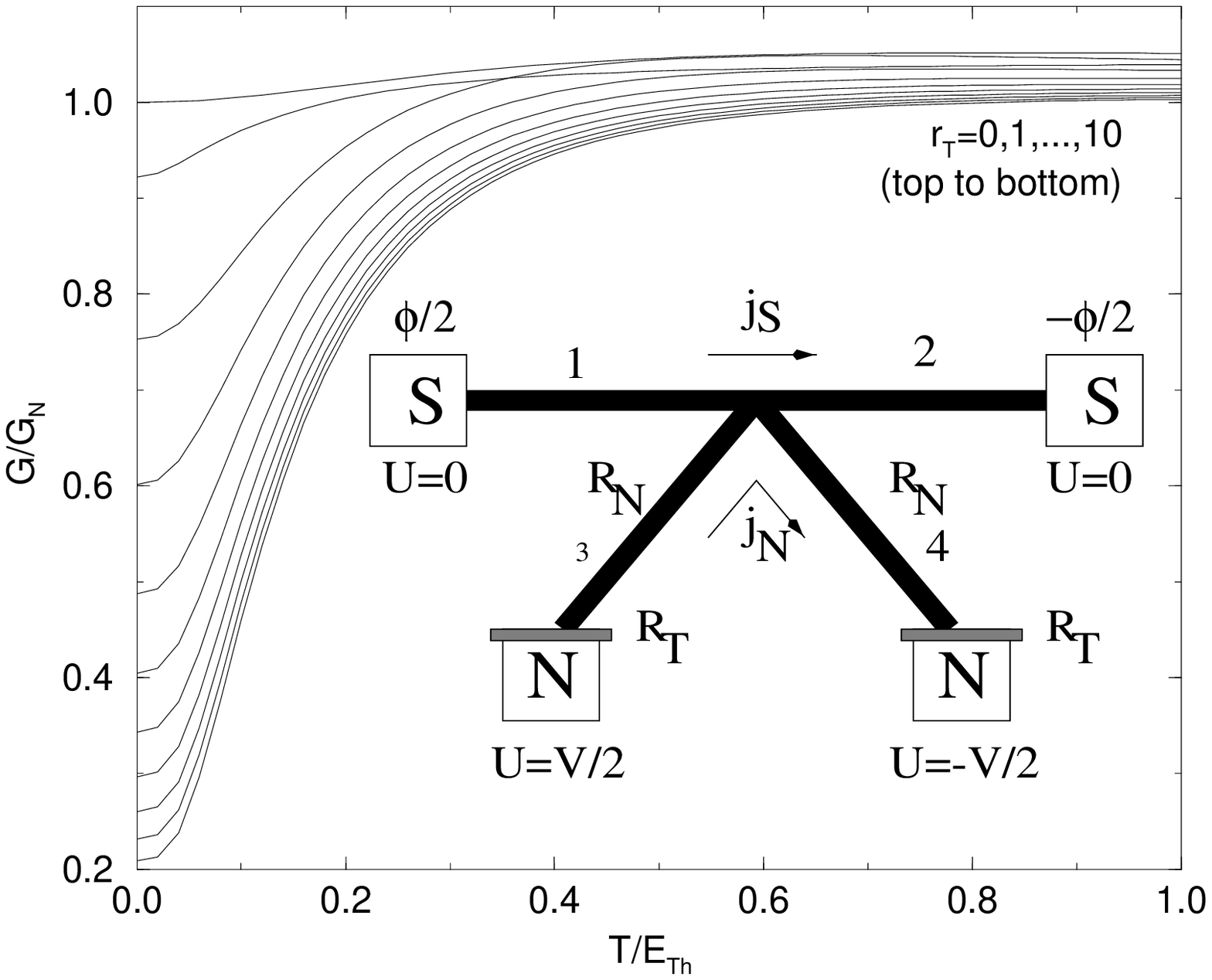,width=70mm}
\caption{Linear conductance as a function of temperature at different 
tunneling resistances within the diffusion/tunneling crossover at $\phi=0$. 
At $\phi=\pi$ we have $G=G_{\rm N}$. The inset shows the structure 
used for calculations. All arms 1--4 are assumed to have the same 
lengths $d$ and cross sections $A$.}
\label{rt1}
\end{minipage}
\hfill
\begin{minipage}{72mm}
\vspace{-10mm}
\psfig{figure=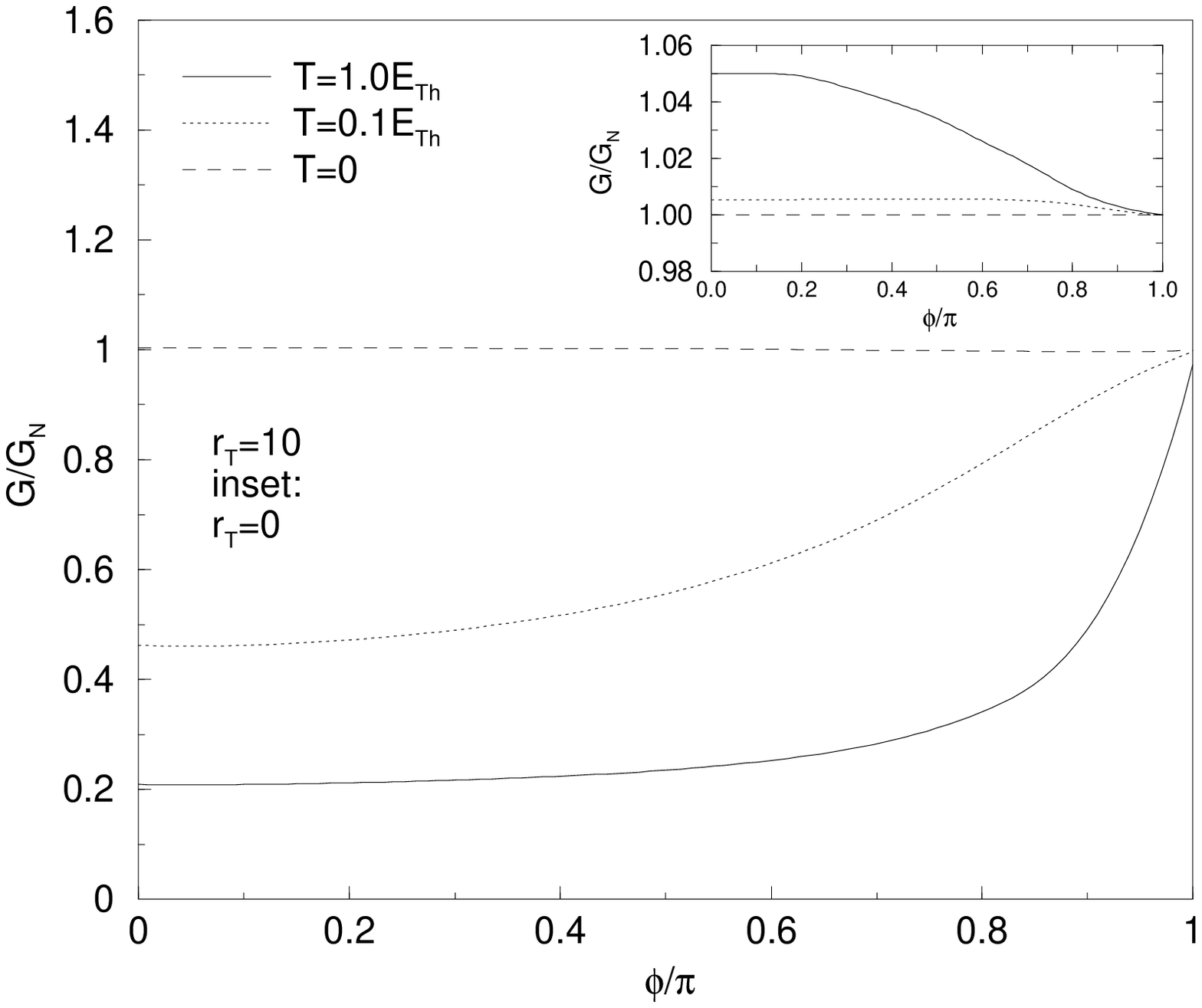,width=70mm}
\vspace{-2mm}
\caption{Linear conductance of the interferometer in
Fig. \ref{rt1} as a function of the phase in the 
tunneling-dominated regime; inset: diffusion-dominated regime.}
\label{rt2}
\end{minipage}
\end{figure}

Following the lines of Ref.~\cite{VZK} with
boundary condition (\ref{GolKupBC}), we can calculate the 
distribution functions and from there the current and
conductance.
The result has
the form (\ref{ZaitsevCond}), however, the  spectral conductance
now accounts for the series of the tunneling resistance and the
diffusive metal
\begin{equation}
g(E)=G_N\frac{1+r_{\rm T}}{M(E)+r_{\rm T}/\nu_{\rm T}(E)}\; .
\label{eq:ohm}
\end{equation}
Here $r_{\rm T}=R_{\rm T}/R_{\rm N}$, where $R_{\rm N}$ is the normal
state resistance of the wire, and $\nu_{\rm T}=\hbox{Re} \{G^{\rm R}\}$
is the reduced DOS at the interface. This result implies that already
a rather low barrier of size $\sim R_{\rm N}$ gives rise to a
tunneling-like structure. In typical experiments, this is of order
$1\dots10\Omega$. In Fig. \ref{rt1} we show the conductance of a
structure relevant for experiments~\cite{ATWZ} demonstrating that at
low $T$ for $r_T>1$ the conductance is reduced compared to the normal
state value.

As we showed in section \ref{ch:oszi}, a magnetic flux through an
Andreev interferometer causes interference effects which influence the
conductance. This interference turns out to be destructive. For the
system without barriers studied in section \ref{ch:oszi}, this
manifests itself in the fact that the conductance oscillations start
from a zero-flux maximum.  If the interferometer contains tunneling
barriers, we expect that the interference effects inside the
interferometer are qualitatively not affected, however the conductance
eq. (\ref{eq:ohm}) is dominated by tunneling into the induced gap.
The destructive interference will weaken induced superconductivity,
therefore lift the gap and consequently increase the conductance. Thus
in this system we also expect magneto-conductance oscillations as
described in section \ref{ch:oszi}, however, they will depart from a
zero-flux minimum.  By combining the theory of the interferometer developed
in section \ref{ch:oszi}
with the conductance formula (\ref{eq:ohm}), 
we can study these structures and
confirm this picture \cite{ATWZ,VolZai}. In Fig. \ref{rt2} we show a
half-period of the conductance oscillations at different temperatures
for a tunneling-dominated system in comparison to the same system
probed through metallic contacts which have a different sign of the
flux-dependence.  This effect has recently also been observed in
experiments \cite{ATWZ}.\\

\subsection{Four-Point conductance}
In the previous sections we argued, that the proximity effect increases the
conductivity of a normal metal wire compared to the normal state. If
the current is probed through tunneling barriers, this increase competes with
the opening of a superconducting gap which can cause the total conductance
to decrease.  
However, a decrease in the conductance was also observed in experiments 
\cite{PetrJETPL93,PetrPRL95}
where the current does not have to pass through tunneling barriers.
This effect {\em cannot} be explained by the non-monotonic conductance described
in section \ref{ch:reswire}, 
because despite its T-dependence, the correction to the normal state
conductance is still non-negative.

In the experiments\cite{PetrJETPL93,PetrPRL95}, the current was probed by
a particular four-probe configuration on a sample of appreciable width.
Hence the current flow cannot be described by simple quasi-1D models as above. 
We will now develop a quasiclassical description of such measurements and 
show how the inhomogenous conductivity of the sample leads to a decrease of the
four-point conductance.

\begin{figure}
\centering
\begin{minipage}{72mm}
\psfig{figure=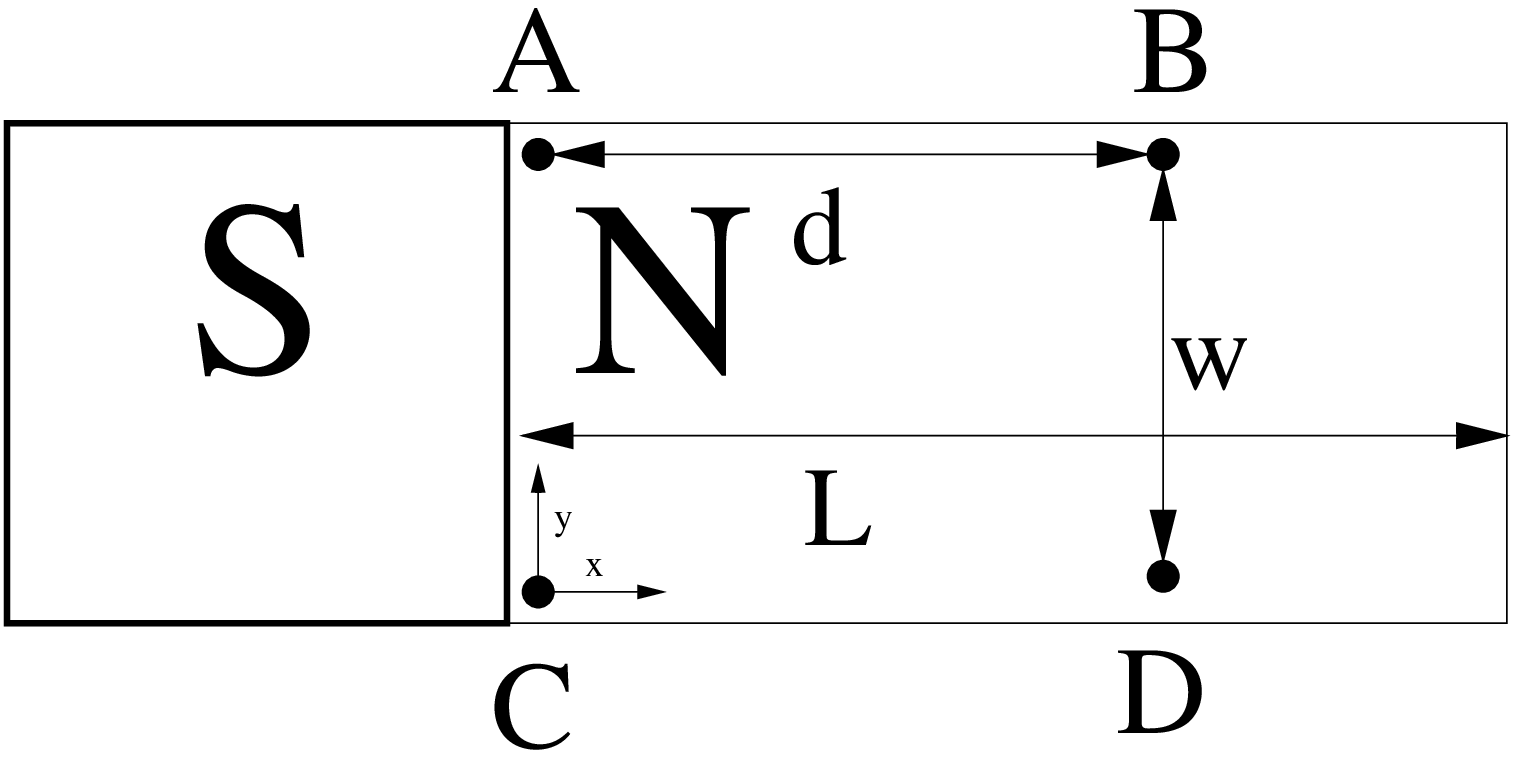,width=60mm}
\caption{A quasi-2D proximity film. The contacts A and B are used
as voltage and C and D as current probes. An alternative setup:
A and C are voltage probes, while the current flows through B and D.}
\label{film}
\end{minipage}
\hfill
\begin{minipage}{72mm}
\psfig{figure=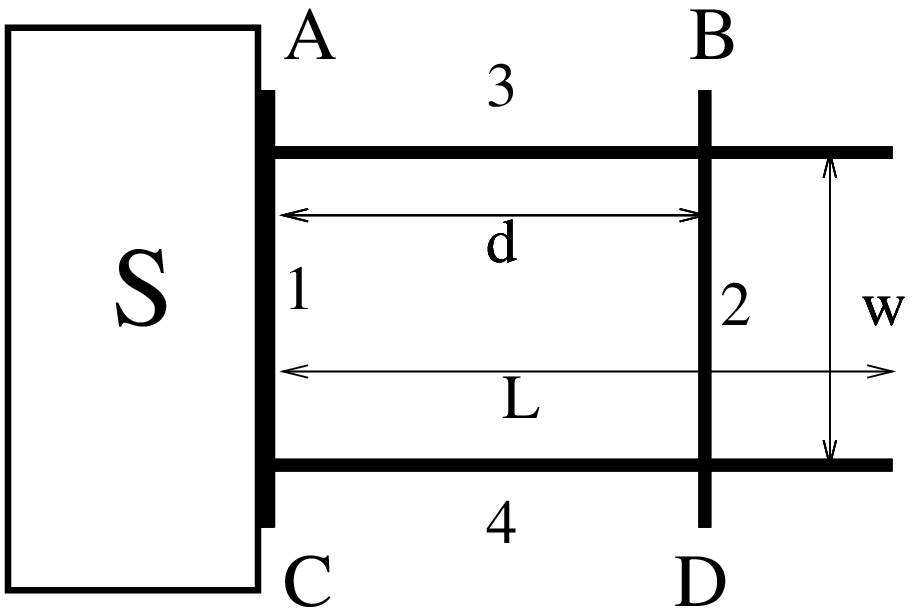,width=60mm}
\caption{A qualitatively equivalent circuit with the probe 
configuration as in Fig. \ref{film}.}
\label{network}
\end{minipage}
\end{figure}

\begin{figure}
\centerline{\psfig{figure=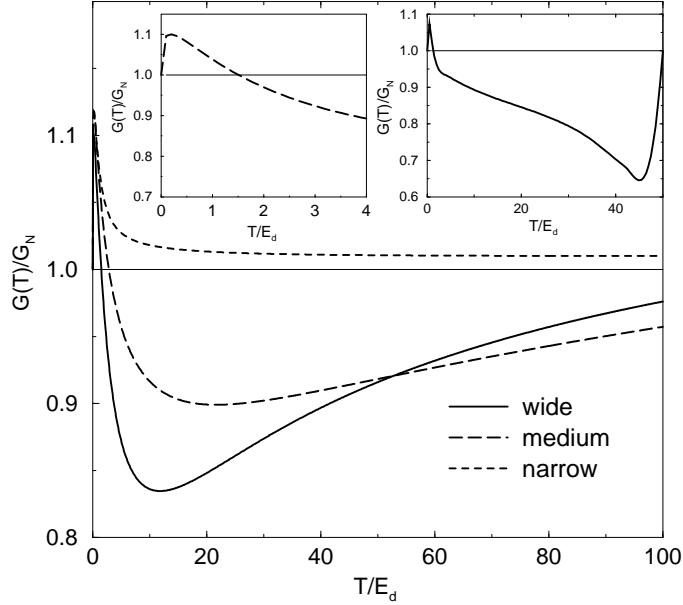,width=95mm}}
\caption{The linear conductance $G(T)$ for films of different widths
 $w/d=0.05,0.5,1.0$ calculated for $d=L$ and $T_{\rm c}=5.7\,10^5E_{\rm Th}$.
 Left inset: The same curve at $w=0.5d$. The $T$-axis is zoomed to
 demonstrate the presence of a usual 1D-type non-monotonic behavior at $T
 \sim E_{\rm Th}$. Right inset: $G(T)$ for a wide film and $T_{\rm
  c}=50E_{\rm Th}$. The amplitude of the negative conductance peak is
 increased due to the effect of a superconducting gap $\Delta (T)$.}
\label{trans2d}
\end{figure}

Let us consider a normal film connected to a superconductor with four point-
like contacts, which lead into measuring reservoirs after a distance
much larger than $l_{\phi}$, see Fig. \ref{film}.
In this configuration, the probes do not influence the spectral
properties, so the solution of the retarded Usadel equation is
analogous to the quasi-one dimensional case with an infinite tunneling
barrier separating the normal reservoir from the sample
\begin{equation}
\partial_x\theta (x=d)=0\; .
\end{equation}
At large energies, $\theta$ is small  and we reproduce
(\ref{HighT}). At low energies, a perturbative expansion gives
\begin{equation}
\theta_E(x)={E\over E_{L}}{x\over L}\left(2-\frac{x}{L}\right)-i\pi/2\; .
\end{equation}
The kinetic equation is now two-dimensional where the current contacts
are represented by source terms
\begin{equation}
\nabla({\cal D}_{\rm T}(r)\nabla f_{\rm
T})=I_E(\delta(r-r_{C})-\delta(r-r_{D}))\; , 
\label{kinetic2D}
\end{equation}
A ``no current flow'' condition at the normal metal edges yields
\begin{equation}
\partial_nf_{\rm T}=0\; .
\label{EdgeCondition}
\end{equation}
A formal solution of Eq.~(\ref{kinetic1}) reads
\begin{equation}
f_{\rm T}(E,r)=I_E({\cal G}_E(r,r_C)-{\cal G}_E(r,r_D))\; ,
\label{formal}
\end{equation}
where ${\cal G}_E=({\cal D}_{\rm T}\nabla + {\cal D}_{\rm T}\nabla
^2)^{-1}$ is the Green's function of the operator
(\ref{kinetic1}). Making use of (\ref{KineticBoundary},
\ref{EdgeCondition}) and (\ref{formal}), and integrating $I_E$ over
energy we obtain the total current $I$ and arrive at the expression
for the differential four-point-conductance $G=dI/dV$:
\begin{equation}
G(V,T)=\int_0^\infty \frac{g(E)}{2T\cosh^2((E-V)/2T)}dE\; ,
\end{equation}
where
\begin{equation}
g(E)=G_{\rm N}\frac{{\cal G}_0^{BC}-{\cal G}_0^{BD}-{\cal
G}_0^{AC}+{\cal G}_0^{AD}}
{{\cal G}_E^{BC}-{\cal G}_E^{BD}-{\cal G}_E^{AC}+{\cal G}_E^{AD}}
\label{g}
\end{equation}
is the spectral conductance.  We introduced the notation ${\cal
G}^{ij}={\cal G}(r_i,r_j)$ and ${\cal G}_0$ is the Green's function of
(\ref{kinetic1}) in the normal state ($M_E(r)=\sigma_{\rm N}$).  The
spectral conductance (\ref{g}) calculated numerically from
Eqs.~(\ref{DiffusionT}), (\ref{kinetic2D}) and (\ref{EdgeCondition})
is presented in Fig.~\ref{trans2d}.

For narrow films the well-known results of quasi-1D calculations
\cite{NazStPRL,GWZ} are qualitatively reproduced: the linear
conductance $G(T)$ exceeds $G_{\rm N}$ at all temperatures, showing a
non-monotonic feature at $T \le E_{d}$ (for simplicity we put $L=d$
here and below).  
s
For broader films $G(T)$ {\it decreases} below the normal-state value
at high temperatures and reaches the minimum at $T \sim 10 E_{\rm
Th}$. At lower $T$ the conductance grows with decreasing $T$, becomes
larger than $G_{\rm N}$ and then decreases again down to
$G(T=0)=G_{\rm N}$ similarly to the 1D case (see the left inset in
Fig.~\ref{trans2d}). The behavior of $g(E)\equiv G(E,T=0)$ as a
function of energy (voltage) is qualitatively identical to that of
$G(T)$, the negative peak at $E \sim 10E_{\rm Th}$ turns out to be
even somewhat deeper.

In order to develop a semi-quantitative understanding of these
results, let us consider a wire network representing the
interconnections between the probes as in fig. \ref{network}. 
A similar equivalent-circuit
model was previously used for a qualitative description of
inhomogeneous superconducting films \cite{Vaglio}. Exploiting the
analogy between $f_{\rm T}$ and the electrical potential in a
conventional circuit (see above and \cite{WZC}), 
Kirchhoff's laws for the spectral
conductances can be derived \cite{GWZ,ZaiNet}. For the present
circuit Sig. \ref{network}, we find (c.f.  \cite{Vaglio})
\begin{equation}
g_{\rm Net}=g_3g_4\sum_{i=1}^4 g_i^{-1}\; ,
\label{KirNet}
\end{equation}
where the $g_i$ are the spectral conductances\cite{VZK,GWZ} of the
wires 1--4 calculated analogous to previous sections. At $T \gg E_{\rm
Th}$ only the wire 1 directly attached to a superconductor acquires
superconducting properties, whereas the proximity effect in the wires
2, 3 and 4 is suppressed. Thus only $g_1$ increases, and $g_{2,3,4}$
remain unaffected. According to Eq.~(\ref{KirNet}) $g_{\rm Net}$
decreases below $G_{\rm N}$. At $T \approx E_{\rm Th}$ the proximity
induced superconducting correlation penetrates into all four wires,
i.e. $g_{2,3,4}$ increase, leading to the increase of $g_{\rm Net}$ above
$G_{\rm N}$. Close to $\Delta$, the conductance correction has a peak
corresponding to the peak in the DOS. If close to $T_{\rm c}$,
$\Delta$ becomes close to $E_{\rm Th}$, the conductance of
interconnection 1 is peaked and the whole effect becomes even
stronger, in agreement to our numerical result.

A semi-quantitative analysis \cite{WZC} shows, that the crossover
energy between excess and deficit conductance is of the order of 
$\hbox{max} (D/d^2,D/W^2)$.  A careful analysis of the experiments
\cite{PetrJETPL93,PetrPRL95} shows that this explains the different
signs of the conductance for different measurements \cite{WZC}.\\

\section{Conclusion}

We outlined the concepts of quasiclassical Green's
functions. They provide a unified view of ballistic and diffusive
systems in equilibrium and nonequilibrium situations. 
We applied the theory to describe  mesoscopic proximity structures. 

We acknowledge useful and stimulating discussions with V.~N. Antonov,
J.~J.~A.  Baselmans, G. Blatter, H. Courtois, M. Devoret, P. Dubos, D.
Est\'eve, A.  Fauch\`ere, M. Fogelstr\"om, A.~V. Galaktionov, A.~A.
Golubov, S.~G. den Hartog, T.~M. Klapwijk, C.~J. Lambert, B. Meyer, B.
M\"uller-Allinger, A. Morpurgo, A.~C. Mota, B.  Pannetier, V.
Petrashov, A. Poenicke, H. Pothier, R. Raimondi, E. Scheer, R.
Seviour, H. Takayanagi, A. Tagliacozzo, A.~F. Volkov, H. Weber, and
B.~J. van Wees. This work was supported by the DFG through
Sonderforschungsbereich 195 and a Graduiertenkolleg.

\label{lastpage}

\end{document}